\documentclass[prd,twocolumn,floatfix,amsmath,nofootinbib,amssymb,floatfix]{revtex4}
\usepackage{graphicx,color,dcolumn,booktabs,bm}
\usepackage{longtable,lscape}
\usepackage{pdfpages}
\usepackage{txfonts}
\usepackage{overpic}
\usepackage{amssymb}
\usepackage{makecell}
\usepackage{indentfirst}
\usepackage{feynmf}   
\usepackage{slashed}  
\usepackage{cases}
\usepackage{color}
\usepackage{multirow}
\usepackage{threeparttable}
\usepackage{epstopdf}
\usepackage{enumerate}
\usepackage{subfigure}
\usepackage{diagbox}
\usepackage{graphicx,color,dcolumn,booktabs,bm}
\usepackage{mathrsfs}
\usepackage{cancel}
\usepackage{float}
\usepackage[colorlinks,
            citecolor=blue,
            anchorcolor=red,
            menucolor=red,
            linkcolor=red,
            filecolor=red,
            runcolor=red,
            urlcolor=blue,
            frenchlinks=red]{hyperref}

\begin{document}
\title{Low-lying singly heavy baryon states based on the rigorous calculation with the relativized quark model}
\author{Zhen-Yu Li$^{1}$}
\email{zhenyvli@163.com }
\author{Guo-Liang Yu$^{2}$}
\email{yuguoliang2011@163.com }
\author{Zhi-Gang Wang$^{2}$ }
\email{zgwang@aliyun.com }
\author{Jian-Zhong Gu$^{3}$ }
\email{gujianzhong2000@aliyun.com }

\affiliation{$^1$ School of Physics and Electronic Science, Guizhou Education University, Guiyang 550018,
China\\$^2$ Department of Mathematics and Physics, North China Electric Power University, Baoding 071003,
China\\$^3$ China Institute of Atomic Energy, Beijing 102413,China}
\date{\today }

\begin{abstract}
In this work, the low-lying $\mathbf{6}_{F}$ singly heavy baryon states with positive-parity are studied in detail in the framework of the relativized quark model by using the improved calculation scheme which has nicely explained the fine structure of the low-lying negative-parity singly heavy baryons. The low-lying spectra of all the singly heavy baryon families obtained in the same framework and calculation scheme are systematically analyzed. The baryon states marked with (mass)$J^{P}$ are obtained by considering the mixing effect rigorously. It is found that the mixing effect in the singly heavy baryons depends on the flavor symmetry of the two light quarks and the baryon parity. The results show that the high-precision calculation reproduces the majority of the assigned experimental masses within the maximum deviation range of 20 MeV and the mean absolute deviation between the calculated masses and the experimental data is only 6.88 MeV. The rigorous calculation achieved by the two-step GEM enables us to analyze the detailed behavior of the various strong interaction components within the baryons with a high-precision. The large amount of data obtained in this work serves as the useful references for related experimental and theoretical researches.

\end{abstract}
\maketitle
\section{Introduction}\label{sec1}
The baryon spectroscopy contains a wealth of information about the strong interaction in the low-energy region, which is crucial for deeply understanding the properties of the non-perturbative Quantum Chromodynamics (QCD)~\cite{F101}. Especially, the singly heavy baryons have attracted great attention as their spectral structure is simpler than that of the light-quark baryons. So far, a large number of heavy baryons have been measured in experiment~\cite{F201,F2021,P2940,F205,Babar06,P2910,P2882,F206,F207,F208,LHCb2002,F210,F209,F211,F212,F213,F4401,F4402,F4403,F203,F204,F214,LHCb25,LHCb26,LHCb262}, which provides abundant experimental data for the theoretical research~\cite{CHY15,P251,F411,Chenhx23,Crede24}. In the new Review of Particle Physics (RPP) by the Particle Data Group (PDG), more than 70 singly heavy baryons have been collected~\cite{F201}. The rudimentary structure of the low-lying singly heavy baryon spectroscopy has already emerged, which provides one of the accessible ways to understand the behavior of the strong interaction dominated by the non-perturbative QCD. Now, how to correctly interpret these experimental data and draw a reliable mass spectrum for these baryons has become one of the hot topics in the current theoretical research.

In systematically analyzing the baryon spectral structure, the quark potential model offers distinct advantages over alternative theoretical approaches~\cite{F306}, including lattice QCD~\cite{MP17,HB20,Chen23}, QCD sum rules~\cite{P21,P22,P221,F315,f17,f17p8,f17p7}, effective field theory~\cite{F304,F303}, chiral perturbation theory~\cite{P23,P24} and Regge trajectories~\cite{P25,CJK24}. Among various quark potential models~\cite{CB15,F403,F405,RNF22,YG18,WKL19,f17p13,f17p10,F601,zhao24,F315p,F4071,f17p14,f17p11}, the relativized quark model (RQM)~\cite{F401,F402}, went through a long and tortuous journey in systematically describing the mass spectra of baryons, during which the theoretical efforts were guided by two central questions: (1) How to understand the spectral structure of baryons? (2) How to calculate the baryon spectra reasonably?

Influenced by the earlier works~\cite{Cor75,Isg77,Fp001,Isg79,Isg78,Isg81}, when the RQM was extended to study the baryons in 1986~\cite{F402}, the authors adhered to describe the light-quark baryons and the singly heavy baryons in a unified framework and adopted the following calculation scheme. By considering the mixing effect between different orbital excited states having the same $J^{P}$ values, the baryon spectrum of $J^{P}$ was obtained in the basis of the $N_{max}$ harmonic-oscillator wave functions, in which the total orbital angular momentum $L$ was used in the calculation but never regarded as a good quantum number. In this calculation scheme, the mixing effect is of great significance for the fact that a real baryon state (or a physical observed state) can only be marked with its spin-parity $J^{P}$.

In this calculation scheme, however, the RQM predicted more excited states than the observed ones in both the light-quark baryon spectra (i.e. the puzzle of the missing states)~\cite{F101,F402} and the recently calculated heavy baryon spectra~\cite{Fp003}. Moreover, due to the difficulty of the three-body problem, some important interaction components in the RQM were omitted in calculation, such as the spin-orbit three-body potentials~\cite{F402,Fp001,zhao24}. In short, the RQM encountered two problems from the very beginning when it was used to study the baryons in 1986, they are the calculation scheme and the high-precision calculation.

On the one hand, for understanding the puzzle of the missing states and finding a better calculation scheme, a possible solution is to freeze some dynamical degrees of freedom~\cite{CPS00}. To this end, one can also use the diquark approximation~\cite{MA93}, although which is not very good in describing the light-quark baryons and still a controversial method~\cite{F101}. Later, the heavy quark symmetry~\cite{Isgur91}, the heavy quark limit~\cite{GG90} and the heavy quark effective theory~\cite{F403,EE90} were put forward one after another, which revealed some important structure properties of heavy baryons. Heavy baryons became one of the breakthroughs in the study of the baryon spectral structure. Quickly, the diquark approximation was successfully applied to analyze the singly heavy baryon spectra~\cite{F405}. It turned out that the predicted states in the diquark approximation are significantly fewer than those in Ref.~\cite{F402} and reproduce most of the experiment data nicely.

The success of the diquark approximation in singly heavy baryons implies two key points: (1) The mixing effect could be temporarily ignored and the total orbital angular momentum $L$ can be approximatively regarded as a good quantum number, even though it is not strictly true in a relativistic theory; (2) The orbital excitation between the two light quarks could be frozen.

Inspired by the diquark approximation, some theoretical works investigated the orbital excitation of singly heavy baryons in the three-quark system~\cite{F403} and found that the excitation energy of the $\lambda$-mode (see Sec.~\ref{sec2}) is commonly lower than that of the $\rho$-mode in a singly heavy baryon~\cite{F403,YO15,Isgur81} and the orbital excitation of singly heavy baryons is mainly dominated by the $\lambda$-mode~\cite{Chen19}, where the orbital excitation between the two light quarks is forbidden just like the key point (2) mentioned above. This might be the reason why the diquark approximation succeeded. Later, it was found that the orbital excitation properties of singly and doubly heavy baryons can be understood naturally by the mechanism of the heavy-quark dominance (HQD)~\cite{Li241}, which also determines the structure of the excitation spectra of these heavy baryons. This might help in understanding the structure of the excitation spectra in singly and doubly heavy baryons.

On the other hand, the Gaussian expansion method (GEM)~\cite{Ka88,Hi03} has been introduced in the calculation of the multi-quark system in recent years~\cite{YG18,YO15,F601,Luo23,Luo25}. This method has the advantage in the high-precision calculation for a given orbital excited state. Especially, the infinitesimally-shifted Gaussian (ISG) basis function method can work out the complicated integration over all angular coordinates. The GEM (ISG) provides an important technical support for the high-precision calculation in the quantum few-body system.

Building upon the above two progresses, the low-lying orbital excited states of singly and doubly heavy baryons were studied rigorously in the genuine three-quark picture, based on the RQM, HQD mechanism, GEM and ISG basis function method~\cite{Li251,Li252}. The used calculation scheme is named as the combination method of the RQM + HQD + GEM(ISG)~\cite{Li252}. It was shown that in the genuine three-quark picture, most of the experimental data can be reproduced nicely and the diquark approximation is not necessary~\cite{Li251}. With this combination method, all the rigorous calculations have been achieved for the orbital excited states and the mean absolute deviation (MAD) between the calculated and measured masses is less than 10 MeV for the singly heavy baryons~\cite{Li251}. Therefore, this combination method offers a better calculation scheme than that proposed in 1986~\cite{F402}. However, in this method, $L$ is still a good quantum number and the mixing effect is not taken into account. So, this combination method~\cite{Li251} is not yet fully developed.

With the improvement in the measurement precision thanks to the advances of the high energy facilities and detectors in recent years, the fine structure in the experimental spectra of the negative-parity singly heavy baryons was revealed and received attention from theoretical research~\cite{Li242}, which requires a more precise theoretical analysis. To meet this requirement, the theoretical (model) calculation should not adopt any approximate methods, because any approximation could result in some unknown systematical errors and unreliable conclusions~\cite{Zhu26}.
To this end, the combination method of the RQM + HQD + GEM(ISG) has to be improved to consider the mixing effect and the fact that $L$ is not a good quantum number indeed, which ultimately returns to the original intention of the earlier work~\cite{F402} as discussed above. For taking into consideration of the mixing effect, however, the complicated calculation poses a great challenge for the theoretical study of heavy baryons.

Actually, the rigorous calculation of some interaction terms in the RQM for the baryon system has always been a bottleneck for nearly 50 years. For instance, the three-body spin-orbit potentials are extremely difficult to handle~\cite{zhao24}. Their contributions disappear automatically in the matrix elements within a single orbital excited state~\cite{Li251}, but necessarily appear in the matrix elements between different orbital excited states (see Appendix). Very recently, this bottleneck has been broken by introducing the two-step GEM, an innovation of the traditional calculation of the GEM, and the spectral fine structure of the negative-parity singly heavy baryons was analyzed nicely with the improved calculation scheme~\cite{Li26}.

At this point, the two problems that the RQM encountered as discussed above have been partially resolved after 40 years of exploration. Meanwhile, the understanding of spectral structure for the heavy baryons has also been further deepened.
This might pave the way for the high-precision spectral analysis of singly heavy baryons. Along this way, reliable mass spectra of low-lying excited singly heavy baryons should be drawn. This is precisely one of the tasks of this work.

The remainder of this paper is organized as follows. In
Sec.~\ref{sec2}, the improved calculation scheme used in this work is introduced, including the wave functions, Jacobi coordinates, Hamiltonian of the RQM, HQD mechanism, GEM and calculation of the mixing effect. The structural properties of all the low-lying singly heavy baryon spectra (especially those of the $\mathbf{6}_{F}$ states with positive-parity), mixing effect, analysis of the experimental data and reliability of the calculation scheme are analyzed in Sec.~\ref{sec3}. And Sec.~\ref{sec4} is reserved for the conclusions.

\section{The improved calculation scheme}\label{sec2}

The improved calculation scheme is different from that used in the 1986 paper~\cite{F402}, which has been gradually established after a long period of exploration as discussed in Sec.~\ref{sec1}. Briefly, the improved calculation scheme is also a combination method which combines together the RQM, HQD, GEM (ISG), mixing effect, two-step GEM and wave function with appropriate symmetry~\cite{Li26}.

1) Wave function

\begin{figure}[htbp]
\centering
\includegraphics[width=8.5cm]{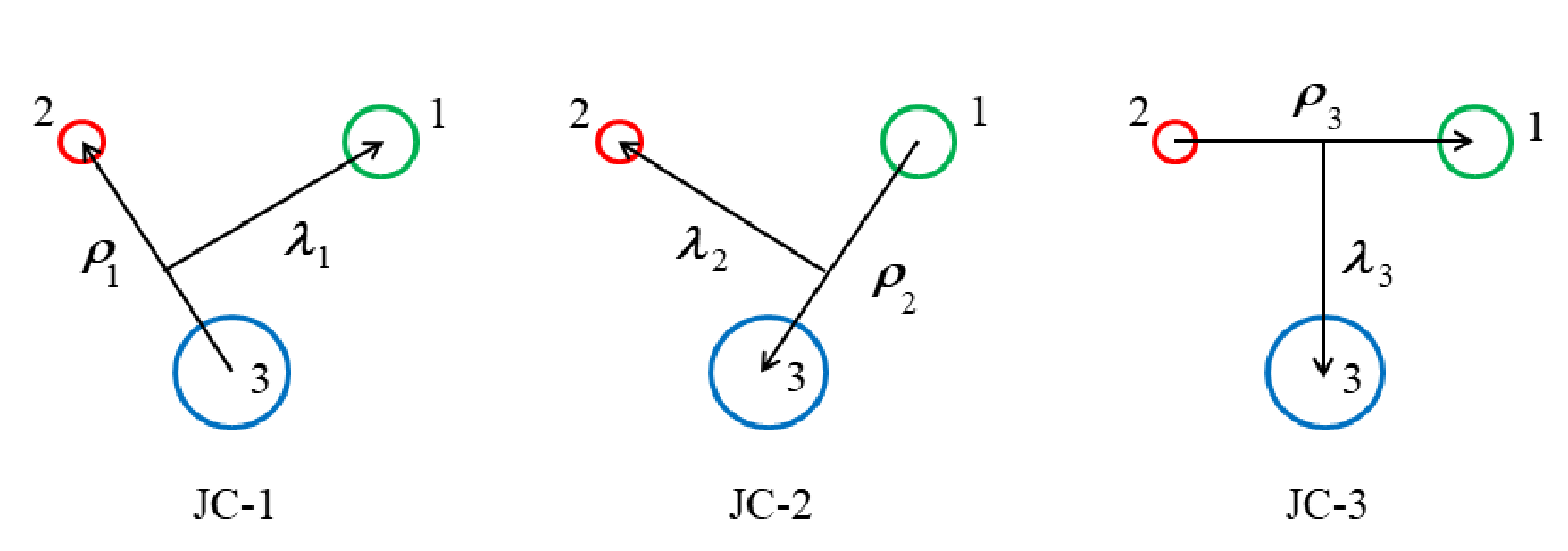}
\caption{There are 3 channels of the Jacobi coordinates for a three-quark system, labeled with $(\boldsymbol\rho_{k}$, $\boldsymbol\lambda_{k})$ ($k$=1, 2, 3). The channel 3 (JC-3) is selected for defining the spin and orbital wave functions in the singly heavy baryon system. All the quarks are numbered for ease of calculation, and the 3rd quark refers specifically to the heavy quark.}
\label{fig1}
\end{figure}

In a singly heavy baryon, the two light quarks are assumed to satisfy the flavor $SU(3)$ symmetry. So, the singly heavy baryon families can be divided into two sectors, i.e., the sextet ($\mathbf{6}_{F}$) of the flavor symmetric states ($\Sigma_{c}$, $\Xi'_{c}$, $\Omega_{c}$, $\Sigma_{b}$, $\Xi'_{b}$ and $\Omega_{b}$) and the anti-triplet ($\mathbf{\bar{3}}_{F}$) of the flavor antisymmetric states ($\Lambda_{c}$, $\Xi_{c}$, $\Lambda_{b}$ and $\Xi_{b}$). For matching with the flavor symmetry, the special Jacobi coordinates JC-3 are selected (see Fig.~\ref{fig1}), with $(\boldsymbol\rho_{3}$, $\boldsymbol\lambda_{3})$ $\equiv$ $(\boldsymbol\rho$, $\boldsymbol\lambda)$ in this work. Then, the spin and orbital wave functions have the following coupling scheme~\cite{F403},
\begin{eqnarray}
|(J^{P})_{j},L\rangle = |\{[(l_{\rho} l_{\lambda} )_{L}(s_{1}s_{2})_{s_{12}}]_{j} s_{3}\}_{J }\rangle,
\label{eq2}
\end{eqnarray}
with parity $P=(-1)^{l_{\rho}+l_{\lambda}}$.
$l_{\rho}$($l_{\lambda}$), $L$ and $s_{12}$ are the quantum numbers of the relative orbital angular momentum $\textbf{\emph{l}}_{\rho}$ ($\textbf{\emph{l}}_{\lambda}$), total orbital angular momentum $\textbf{\emph{L}}$ and total spin of the light-quark pair $\mathbf{s}_{12}$, respectively. $j$ denotes the quantum number of the coupled angular momentum of $\textbf{\emph{L}}$ and $\textbf{s}_{12}$. Here, $(-1)^{l_{\rho}+s_{12}}=-1$ should be guaranteed for the $\mathbf{6}_{F}$ sector due to the total antisymmetry of the wave function of the two light quarks, but $(-1)^{l_{\rho}+s_{12}}=1$ for the $\mathbf{\bar{3}}_{F}$ sector.
Simply, the orbital excited state is labeled with $(l_{\rho},l_{\lambda})L(J^{P})_{j}$.
$(l_{\rho},l_{\lambda})L$ denotes the orbital excitation mode. It is commonly known as the $\lambda$-mode if $l_{\rho}=0$, or the $\rho$-mode if $l_{\lambda}=0$ in the literature~\cite{F101}.

2) RQM

The used quark potential model in this work is the relativized quark model (RQM) proposed in the 1980s~\cite{F401,F402}. The RQM can be recognized as a precise potential model since it takes into account more physical factors concerning QCD and the relativistic correction than other quark models and has been confirmed to be reliable in describing low-lying heavy baryons~\cite{Li251}.

In the RQM, the Hamiltonian reads
\begin{eqnarray}\label{e1}
\notag
\hat{H} &&=\hat{H}_{0}+\hat{H}^{conf}+\hat{H}^{hyp}+\hat{H}^{so}\\
&&=\sum_{i=1}^{3}\sqrt{p_{i}^{2}+m_{i}^{2}}+\sum _{i<j}(\hat{H}^{conf}_{ij}+\hat{H}^{hyp}_{ij}+\hat{H}^{so}_{ij}).
\label{eq3}
\end{eqnarray}
The confinement term $\hat{H}^{conf}_{ij}$ includes a modified one-gluon-exchange potential and a smeared linear confinement potential. The hyperfine interaction $\hat{H}^{hyp}_{ij}$ consists of the tensor term $\hat{H}^{tens}_{ij}$ and the contact term $\hat{H}^{cont}_{ij}$. And the spin-orbit interaction $\hat{H}^{so}_{ij}$ can be divided into the color-magnetic term $\hat{H}^{so(v)}_{ij}$ and the Thomas-precession term $\hat{H}^{so(s)}_{ij}$. $\hat{H}^{tens}_{ij}$, $\hat{H}^{cont}_{ij}$, $\hat{H}^{so(v)}_{ij}$ and $\hat{H}^{so(s)}_{ij}$ belong to the spin-dependent interactions. It has been confirmed that $\hat{H}^{mode}=\hat{H}_{0}+\hat{H}^{conf}$ determines the energy levels of the orbital excitation mode $(l_{\rho},l_{\lambda})L$. While, these spin-dependent terms contribute to the energy level shifts and splittings~\cite{Li251}.

In the practical calculations, after the Jacobi coordinate transformation, the spin-orbit interaction $\hat{H}^{so}_{ij}$ is decomposed into the two-body term $\hat{H}^{so-2B}_{ij}$ and the three-body term $\hat{H}^{so-3B}_{ij}$. Such a decomposition helps to analyze their contributions to the mixing effect~\cite{Li26}.

3) HQD

The mechanism of the heavy-quark dominance (HQD) can be used to determine the practical orbital excited states of singly heavy baryons. For the $L$-wave excitation with $\textbf{\emph{L}}$=$\textbf{\emph{l}}_{\rho}$+$\textbf{\emph{l}}_{\lambda}$, there are an infinite number of orbital excitation modes.
Taking $L=1$ as an example, the excitation modes $(l_{\rho},l_{\lambda})L$ are $(1,0)P$, $(0,1)P$, $(1,1)P$, $(2,1)P$, $(1,2)P$, $(2,2)P$ and so on. We assume that the excitation mode with the lowest energy is most stable and has the greatest probability of being observed experimentally, which dominates the structure of the excitation spectrum. This assumption is summarized as the HQD approximation (or the HQD mechanism)~\cite{Li241}.

As the consequence of the HQD mechanism, the orbital excited states of low-lying singly heavy baryons mainly come from the $\lambda$-modes. But for the $P$-wave orbital excitations of the charm baryons in the $\mathbf{6}_{F}$ sector, the HQD mechanism is slightly broken because the mass of $c$ quark is not heavy enough, where both the $\lambda$-mode $(0,1)P$ and the $\rho$-mode $(1,0)P$ appear in their $P$-wave states. This assertion has been proven to be correct by the accurate analysis of the experimental data related to the fine structure~\cite{Li26}.
In this work, the $S$-, $P$- and $D$-wave states together with their radial excitations of the low-lying singly heavy baryons are investigated systematically.

4) GEM

The Gaussian basis functions $|(nlm)^{G} \rangle$ form a finite-dimensional Hilbert space with $n$= 1$-$ $n_{max}$ and any orbital quantum state with $\{l$, $m\}$ can be expanded in such a space. We introduce the generalized Gaussian basis functions $|(\tilde{n},\alpha)^{G}\rangle$, in which $|\alpha\rangle$ denotes the spin and orbital quantum state $|(l_{\rho},l_{\lambda})L(J^{P})_{j}\rangle$.
In the generalized Gaussian basis functions, the Hamiltonian $\hat{H}$ is represented as a matrix with $H_{nn'}$ = $\langle(n,\alpha)^{G}|\hat{H}|(n',\alpha)^{G}\rangle$. Its solution belongs to a generalized matrix eigenvalue problem
\begin{eqnarray}
\begin{aligned}
\sum^{n_{max}}_{n'=1}(H_{nn'}-EN_{nn'})C_{n'}=0.
\label{eq4}
\end{aligned}
\end{eqnarray}
Here, $N_{nn'}$ comes from the non-orthogonality of the basis functions. Then, the eigenvalue $E_{n\alpha}$ and the eigenvector $\{C_{n'}\}_{n\alpha}$ ($n'=1-n_{max}$) are obtained for the $|(n,\alpha)\rangle \equiv |(l_{\rho},l_{\lambda})nL(J^{P})_{j}\rangle$ state.
These calculations were performed rigorously in our previous work~\cite{Li251}.
Some important issues of the calculation were also extensively demonstrated, such as the improvement of the Hamiltonian parameters~\cite{Yu23}, modification of the Gaussian size parameters~\cite{Li23}, proof of the convergence~\cite{Yu23,Li23} and reliability of the calculation~\cite{Li251,Li252}.

5) Mixing effect

The above orbital excited states with the same $J^{P}$ values form a subspace. The physical observed states (or the baryon states) can be obtained by diagonalizing the Hamiltonian matrix in this $J^{P}$ subspace. The baryon states are the eigenstates which result from the superposition (mixing) of the orbital excited states (the basis states). These basis states are mixed by the diagonalization, which causes the energy level shifts. So, the mixing effect essentially comes from the diagonalization.

In Ref.~\cite{Li26}, the negative-parity $\mathbf{6}_{F}$ singly heavy baryon states were calculated, where the mixing effect occurs within the 1$P$-wave states. In this work, we focus on calculating the positive-parity $\mathbf{6}_{F}$ singly heavy baryon states, which come from the mixing of the $1S$-, $2S$-, $3S$- and $1D$-wave states of low-lying singly heavy baryons. For overcoming the difficulty in calculating the Hamiltonian matrix elements, the two-step GEM has been adopted, which is an efficient and powerful method to work out the very complicated matrix elements in the Gaussian basis functions~\cite{Li26}.

\begin{figure*}[p]
\centering
\includegraphics[trim=1cm 7cm 6cm 3cm, angle=270]{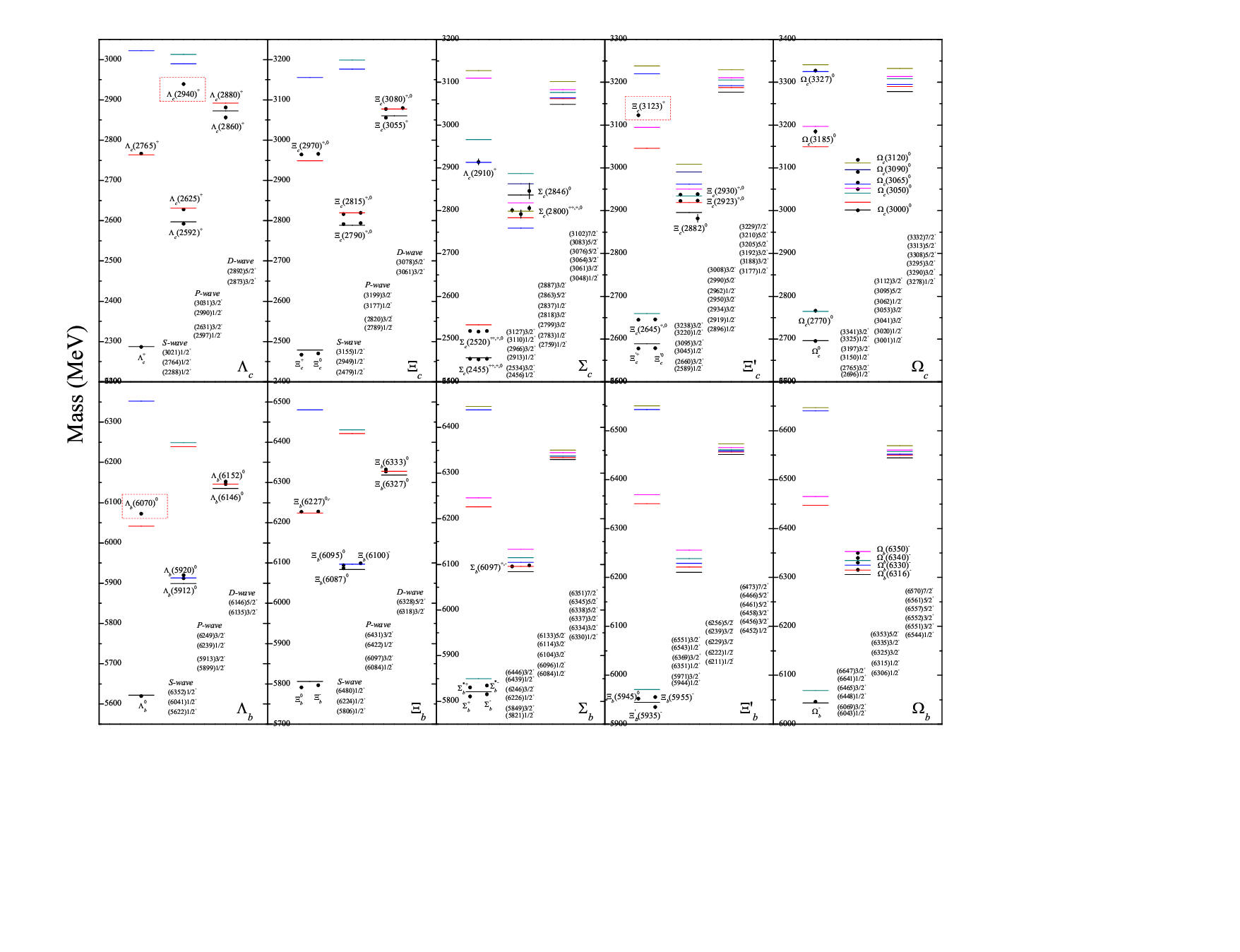}
\caption{Calculated spectra of the singly heavy baryons and the relevant experimental data~\cite{F201,F209,LHCb25,LHCb26}. The predicted energy levels are indicated by the horizontal lines. The baryon states are labeled with (mass)$J^{P}$ and arranged in the lower right corner in each panel, being divided into three groups. The baryons $\Lambda_{c}(2940)^{+}$, $\Lambda_{b}(6070)^{0}$ and $\Xi_{c}(3123)^{+}$ are marked with the boxes, which can not be reasonably assigned.}
\label{fig2}
\end{figure*}

\section{Results and discussion}\label{sec3}

In this work, we have studied the detailed contribution of each term in the Hamiltonian to the energy levels (and their evolution) for the 1$S$- and 1$D$-wave states in each singly heavy baryon family of the $\mathbf{6}_{F}$ sector (see Tables~\ref{tb1} and~\ref{tb2} of the Appendix), and also presented their contributions to the mixing effect in each $J^{P}$ subspace (see Tables~\ref{tb3} and~\ref{tb4} of the Appendix). Meanwhile, the Hamiltonian matrices in the $J^{P}$ subspaces and their eigenvalues and eigenvectors are shown in Tables~\ref{tb5} and~\ref{tb6}. Detailed data on the energy level shifts caused by configuration mixing and the competition among various interactions in the mixing effect are presented in the Tables~\ref{tb7},~\ref{tb8} and~\ref{tb9}. The calculated mass spectra (the eigenvalues) are presented in Fig.~\ref{fig2}, for comparison, the relevant measured masses are also shown there. Finally, the deviations of the calculated masses of the 74 singly heavy baryons from the measured ones and their plots with different $nL$-wave state groups are displayed in Table~\ref{tb10} and~\ref{tb11}, respectively. From the calculation results, we can make the following detailed analysis regarding the low-lying single heavy baryons.

1) The $1S$-wave states

In general, $H^{mode}$ determines the energy levels of the excitation modes as shown in Tables~\ref{tb1} and~\ref{tb2}. These tables also show that, for the $1S$-wave states, the contribution to the energy level shift and splitting comes solely from the contact terms $H^{cont}$. $H^{cont}_{12}$ causes a large energy level shift. While, $H^{cont}_{23}$ and $H^{cont}_{31}$  jointly lead to a big energy level splitting. The energy level shifts (or the splittings) do not make much difference for the $\Sigma_{b}$, $\Xi'_{b}$ and $\Omega_{b}$ baryons. The situation is similar for the charm baryons as well, but the splitting magnitude is much larger than that of the bottom baryons. This conforms to the properties of the heavy quark limit~\cite{F403}.

2) The $1D$-wave states

For the $1D$-wave states, $H^{cont}_{12}$ still causes a big energy level shift. However, $H^{cont}_{23}$ and $H^{cont}_{31}$  jointly lead to a small energy level splitting. In this case, the contribution to the energy level splitting mainly comes from the spin-orbital interactions, even though $H^{so(v)}$ and $H^{so(s)}$ partially cancel each other out. In addition, the tensor terms cause a very small energy level splitting, very similar to the contact terms. Compared to the 1$S$-wave states, the splitting magnitude in the $1D$-wave states is significantly reduced. Just like the case of the $1S$-wave states, the splitting of the $1D$-wave states is also consistent with the heavy quark limit.

3) The mixing effect

In Tables~\ref{tb3} and~\ref{tb4}, the most obvious feature is that those matrix elements of $\hat{O}_{ij}$ with $\{i,j\}$ = $\{1,2\}$ in (or between) the $S$- and $D$-wave states vanish, which is the result of the coordinate transformation and the properties of the $\lambda$-mode. In Ref.~\cite{Li26}, the situation is different from the above mentioned since the $\rho$-mode is also involved in those $P$-wave states in the charm baryon families.
Tables~\ref{tb3} and~\ref{tb4} show that the mixing effects (the off-diagonal matrix elements) between the $S$- and $D$-wave states only come from the tensor terms, and their matrix elements are commonly small. For the matrix elements between the different $D$-wave states with the same $J^{P}$, i.e., $\langle4'|\hat{O}|5'\rangle$ and $\langle1''|\hat{O}|2''\rangle$, all the matrix elements of the spin-dependent terms partially cancel each other out, resulting in a small contribution to the mixing effect. All of the spin-dependent terms compete with each other in a complex manner, and neither can be ignored. Among them, the contribution of the three-body spin-orbit terms $\hat{H}^{so-3B}_{ij}$ is also significant.

In these results of the mixing effect, the heavy quark limit remains valid, i.e., the more massive the heavy-quark is, the smaller the mixing effect becomes.
As is shown in Tables~\ref{tb5} and~\ref{tb6}, by taking into account the mixing effect, the energy level shift is very small, which is different from that in the case of the negative-parity baryons~\cite{Li26}. This difference leads to the following interesting structural feature for the excited energy levels of the singly heavy baryons.
For the low-lying $\mathbf{6}_{F}$ singly heavy baryons with positive-parity, the mixing effect can be roughly ignored and $L$ is approximately a good quantum number. But in the case of the negative-parity, it is not true. This might be the reason why the diquark approximation and the HQD mechanism can achieve certain success in describing the low-lying singly heavy baryons, but they cannot explain the fine structure reasonably.

To illustrate the energy level shifts caused by the mixing effect, we have fabricated Tables~\ref{tb7}, \ref{tb8} and~\ref{tb9}. One can find that the energy level shifts $\Delta M$ in Table~\ref{tb7} are on the order of 0.1 MeV when the configuration mixing occurs between the $nS$- and $1D$-wave states, but they are on the order of 1 MeV when the configuration mixing occurs within the $1D$-wave states. The situation in Table~\ref{tb8} is similar to that in Table~\ref{tb7}, but the energy level shifts in Table~\ref{tb8} are roughly one order of magnitude lower than those in Table~\ref{tb7}. While, for the negative-parity heavy baryons as shown in Table~\ref{tb9}, the shifts reach a scale ranging from 1 to 10 MeV, where the configuration mixing occurs within the $1P$-wave states. So, we can conclude that the mixing effect is generally small between the different $nL$-wave state groups, but very significant within the same $nL$-wave state group.

In Tables~\ref{tb7}, \ref{tb8} and~\ref{tb9}, one can also see the contribution of each component of the interaction to the mixing effect. In most cases, the partial cancellation always occurs between the matrix elements of $\hat{H}^{tens}$ and $\hat{H}^{cont}$, of $\hat{H}^{so(v)-2B}$ and $\hat{H}^{so(s)-2B}$ and of $\hat{H}^{so(v)-3B}$ and $\hat{H}^{so(s)-3B}$, respectively. This phenomenon has been mentioned in the early study~\cite{F402}. In this work, it is quantitatively described.

4) The spectral structure

The mass spectra of the low-lying singly heavy baryons are presented in Fig.~\ref{fig2}, where for a comparison, the spectra of the $\mathbf{\bar{3}}_{F}$ baryons from Ref.~\cite{Li251} and those of the negative-parity $\mathbf{6}_{F}$ baryons from Ref.~\cite{Li26} are also presented. In Fig.~\ref{fig2}, the energy levels are the baryon states and marked with (mass)$J^{P}$ instead of the orbital excited states $(l_{\rho},l_{\lambda})nL(J^{P})_{j}$ as presented in~\cite{Li251}.
In fact, there is no such a mixing effect in the spectra of these $\mathbf{\bar{3}}_{F}$ baryons, where the orbital angular momentum $L$ is an observable. So, the $S$-, $P$- and $D$-wave states are arranged separately in each panel as shown in Fig.~\ref{fig2}. Given that the mixing effect is not so significant for the positive-parity $\mathbf{6}_{F}$ baryons and only occurs within the $P$-wave states for the negative-parity $\mathbf{6}_{F}$ baryons, their spectral structure can also be arranged in the manner of the $\mathbf{\bar{3}}_{F}$ baryons, but the orbital angular momentum $L$ is generally no longer an observable for the $\mathbf{6}_{F}$ baryons. Such an arrangement of the mass spectra can clearly demonstrate the fine structure of the negative-parity baryons and indicate which state contributes most to the energy levels of the positive-parity baryons as well.

5) The analysis of the experimental data

The available experimental data of the singly heavy baryons, along with their measured errors, are all plotted in Fig.~\ref{fig2}. Their assignment can be seen in Table~\ref{tb10}. Most of these baryons have been analyzed in many theoretical works. We have compared our results with those from other relevant theoretical works in Refs.~\cite{Yu23,Li23,Li242,Li251}. In this work, we focus on the precision comparison between the experimental data and theoretical results. As is shown in Fig.~\ref{fig2} and Table~\ref{tb10}, most of the experimental data are close to the theoretical values. There is a nice match between the experimental data and the predicted values. Table~\ref{tb10} provides the important information for the confirmation of the $J^{P}$ values for most of the baryons apart from the singly baryons $\Lambda_{c}(2940)^{+}$~\cite{Babar06,P2940}, $\Lambda_{b}(6070)^{0}$~\cite{LHCb2002} and $\Xi_{c}(3123)^{+}$~\cite{F205}.
It should be noted that in Table~\ref{tb10} most of the well-established baryons from the experiments can be well reproduced by the our calculation. For the other baryons, their assignments in Table~\ref{tb10} only indicate that they are the ideal candidates of the predicted baryon states in this work. And the following analyses are based solely on the assignments in Table~\ref{tb10}.

In this work, $\Lambda_{c}(2910)^{+}$ is assigned as the predicted $\Sigma_{c}(2913)\frac{1}{2}^{+}$ due to their close mass values. In fact, there is indeed a possibility that $\Lambda_{c}(2910)^{+}$ belongs to the $\Sigma_{c}$ baryon family experimentally~\cite{P2910}. The two bottom baryons $\Xi_{b}(6227)^{0,-}$ are assigned as the $\Xi_{b}(6224)\frac{1}{2}^{+}$  state by assuming they are positive-parity baryons. If they are negative-parity baryons~\cite{GLS19}, they should belong to the $\Xi_{b}'(6229)\frac{3}{2}^{-}$ state.

6) Three-body spin-orbital force and tensor force

These two interactions are commonly ignored in the baryon system calculations, because they are too tough to calculate. Now, we can accurately calculate their matrix elements with the help of the two-step GEM. The results show clearly how they contribute to the energy level of each orbital excited state, compete with the other interactions and determine the mixing effect together with the other interactions. In fact, the three-body spin-orbital force $H^{so-3B}$ has no contribution to the energy level of each orbital excited state (the basis state)~\cite{Li251}. The tensor force contributes to the energy level slightly, with the maximum being less than 1 MeV as shown in Tables~\ref{tb1} and~\ref{tb2}. As to the contribution to the mixing effect, however, these two interactions become indispensable. As is shown in Tables~\ref{tb3} and~\ref{tb4}, the tensor force is the sole contributor to the mixing effect between the $S$- and $D$-wave states. Meanwhile, between the different $D$-wave states with the same $J^{P}$, i.e., in $\langle4'|\hat{O}|5'\rangle$ and $\langle1''|\hat{O}|2''\rangle$, $H^{so-3B}$ and $H^{tens}$ play more important roles, even though the overall mixing effect is not so significant in the positive-parity $\mathbf{6}_{F}$ baryons. In the negative-parity $\mathbf{6}_{F}$ baryons, the contributions of $H^{so-3B}$ and $H^{tens}$ to the mixing effect are more important, which leads to the unpredictable results~\cite{Li251}.

More details of their contributions to the mixing effect can also be found in Tables~\ref{tb7}, \ref{tb8} and~\ref{tb9}. In short, the three-body spin-orbital force and tensor force make (almost) no contribution to the diagonal mass splittings (or the energy levels of the single orbitally excited states), but play an important role in the off-diagonal configuration mixing (or the mixing effect).

7) Two-step GEM

The main advantage of the two-step GEM is that it helps us achieve the rigorous calculation for all the $H_{23}$ and $H_{31}$ terms in the JC-3 Jacobi coordinates $(\boldsymbol\rho_{3}$, $\boldsymbol\lambda_{3})$, which was once an impossible task, especially for $H^{so-3B}_{23(31)}$ and $H^{tens}_{23(31)}$ terms. This completely removes the reliance on all kinds of approximations in the calculation of complicated matrix elements, such as the reliance on the diquark approximation. As shown in Tables~\ref{tb1},~\ref{tb3} and~\ref{tb4}, $H_{12}$ always vanishes in many cases, while $H_{23}$ and $H_{31}$ play important roles. In many theoretical studies, for instance, Refs.~\cite{Li242,Yu23,Li23}, the diquark approximation was adopted to replace $H_{23(31)}$ with $H_{d-Q}$ (the interaction between the heavy quark and the light-diquark). This resulted in some unreliable conclusions and ultimately failed to correctly describe the fine structure of the negative-parity singly heavy baryons~\cite{Li26}.

8) Mean absolute deviation and model uncertainty

We collect 74 singly heavy baryon data and obtain the mean absolute deviation (MAD) $(\sum_{i=1}^{n}|M_{cal.}-M_{exp.}|_{i})/n$ being 6.88 MeV as shown in Table~\ref{tb10}, in which more data analysis information can be found in the caption such as the data deviation statistics, the root mean square deviation, the sample median and the median absolute deviation. The baryons $\Lambda_{c}(2940)^{+}$, $\Lambda_{b}(6070)^{0}$ and $\Xi_{c}(3123)^{+}$ are not included in this statistics, because they can not be reasonably assigned to the predicted states. In Ref.~\cite{Li251}, the MAD is about 9.12 MeV, where the mixing effect is ignored. In the two works, the Hamiltonians used are the same, regardless of whether the mixing effect is taken into account or not. The sole reason for the improvement of the MAD is the mixing effect. So, the mixing effect is important.
In fact, without the mixing effect, the fine structure of the negative-parity $\mathbf{6}_{F}$ singly heavy baryons can not be explained reasonably~\cite{Li26}.

The theoretical uncertainty is estimated as follows. For the GEM size parameters, the final results are generally stable and independent of these parameters within an approximate complete set in a sufficiently large space~\cite{Hi03}. In practice, we choose a smaller space with the selected GEM size parameter values $\{r_{1}=0.18$ GeV$^{-1}$, $r_{max}=15$ GeV$^{-1}$, $n_{max}=10\}$ in calculations for the computational efficiency~\cite{Li26}. Our calculation shows that the estimated uncertainty caused by the GEM size parameters should be within 2 MeV. For the input parameters of the RQM, it is found that the orbital excitation energy levels decrease significantly as the mass value of the 3rd quark (see Fig.~\ref{fig1}) increases in some range of the light quark mass. For example (see the Fig.4 in Ref.~\cite{Li241}), in the light quark mass range of $0.2\sim0.5$ GeV, the orbital excitation energy levels vary by about $30\sim100$ MeV. While, in the bottom quark mass range of $4.5\sim5.0$ GeV, most of the energy levels vary by about $0\sim20$ MeV. So, the orbital excitation energy levels are generally sensitive to the quark masses. In actual calculations, however, these input parameters are fixed. In Ref.~\cite{F401}, the uncertainty caused by the model deficiency, especially the restriction of the Fock space and the rather schematic relativistic correction was analyzed. It was believed by the authors that the model uncertainty for heavy quark systems should be around 10 MeV. It is expected that for low-lying states of heavy quark systems the model uncertainty could be even smaller. Table~\ref{tb10} shows that the root mean square deviation (RMSD) $\sqrt{\frac{1}{n}\sum_{i=1}^{n}(M_{cal.}-M_{exp.})_{i}^{2}}$ is 8.71 MeV. If we regard this RMSD as the total uncertainty containing the model uncertainty, numerical calculation uncertainty and experimental error, the estimated model uncertainty for low-lying states of heavy quark systems could be less than 10 MeV in this work.

9) Systematic deviation

From the spectral structure in Fig.~\ref{fig2}, one can find the systematic deviation between the data and the calculated results in the $\Xi'_{b(c)}$ and $\Sigma_{b(c)}$ families. While in the $\Omega_{b(c)}$ families, the deviation is not so obvious. This might be related to the fact that the $\Omega_{b(c)}$ baryons contain $s$ quarks instead of $u$ and $d$ quarks, and $s$ quark is more massive than $u$ and $d$ quarks. In the $\Lambda_{b}$ family, the predicted excited energy levels are generally lower than the data. Especially, the deviation of $\Lambda_{b}(6070)^{0}$ is quite significant, while $\Lambda_{b}(6070)^{0}$ has been well established in experiment~\cite{F201}. In the $\Lambda_{c}$ family, the predicted energy levels of the $2P$- and $1D$-wave states are relatively higher.

In order to more clearly observe the systematic deviations, we investigated the residual ($M_{cal.}$-$M_{exp.}$) plots of the 74 baryons selected in Table~\ref{tb10}, which are grouped with $nL$-wave states for different families as shown in Table~\ref{tb11}. For most of the baryon states dominated by the $1S$-wave states, the predicted mass values are greater than the experimental ones. But for the other groups, the situation is the opposite. We guess the main reason might come from the input parameters of the linear confinement potential. As is shown in Table~\ref{tb11}, there are no correlations of the residuals with the factors such as the heavy-quark flavor, number of strange quarks, radial quantum number and orbital angular momentum.

All the above suggests that there may be more than one reason for the deviation in the calculation such as the relativistic correction in the RQM, the size parameters in the GEM and the modification of the linear confinement potential. If the systematic deviation is corrected, the possible $J^{P}$ values for some baryons may be changed as follows: $\Sigma_{b}(6097)$ $\rightarrow$ $\frac{3}{2}^{-}$ and $\Xi_{c}(2923)$ $\rightarrow$ $\frac{3}{2}^{-}$. To reduce the systematic deviation and improve the high-precision calculation, more efforts need to be done.

10) Existing problems and possible improvements

The calculation in this work uses a set of fixed input parameters (see the Table 2 in Ref.~\cite{Li251}), which has been actually used without any changes after it was determined in Refs.~\cite{Yu23,Li23}. In recent years, we have been continuously improving the calculation scheme. Now, the calculation scheme has been improved by taking into account the mixing effect rigorously. However, Fig.~\ref{fig2} shows that the baryons $\Lambda_{c}(2940)^{+}$, $\Lambda_{b}(6070)^{0}$ and $\Xi_{c}(3123)^{+}$ can not be reasonably assigned although the $J^{P}$ values of $\Lambda_{c}(2940)^{+}$ were identified as $\frac{3}{2}^{-}$ and the identification of $\Xi_{c}(3123)^{+}$ remains an open question~\cite{F201}. Meanwhile, the systematic deviation in the $\Xi'_{b(c)}$ and $\Sigma_{b(c)}$ families remains obvious. More details of the systematic deviation are presented in Table~\ref{tb11}. This implies that the further improvements must be considered in other aspects, such as improving the parameters of the RQM, improving the size parameters in the GEM and improving the relativistic correction. In addition, most of the $\rho$-modes are excluded under the HQD mechanism in our studies of the low-lying singly heavy baryons. For the high-lying states, is the HQD still to be used? And how should the $\rho$-mode be handled? These are all interesting questions. Very recently, a new excited $\Sigma_{c}(3200)^{0}$ state was reported by the LHCb collaboration~\cite{LHCb262}. Our calculation indicates that the measured mass reaches the energy region of the $2P$-wave states and is beyond the scope of this work. This new state expands the $\Sigma_{c}$ spectrum and is an interesting topic of research.

\section{Conclusions}\label{sec4}

Finding an effective calculation scheme and overcoming the difficulty in the high-precision calculation have been troubling the RQM for 40 years. After a long time exploration, a significant progress on the two issues has been made in the study of the singly heavy baryons. The improved calculation scheme (RQM + HQD + GEM (ISG) + two-step GEM + mixing effect + wave functions with appropriate symmetry) has been proposed very recently in order to analyze the fine structure of the negative-parity singly heavy baryons with a high-precision. In this work, this improved calculation scheme is used to study the positive-parity $\mathbf{6}_{F}$ singly heavy baryons and the complete mass spectra of the low-lying singly heavy baryons are presented. Based on that, we have analyzed many details of the strong interaction dominated by the non-perturbative QCD .

The high-precision calculation results reveal the contribution of each Hamiltonian term to the energy levels (and their evolution) for the 1$S$- and 1$D$-wave states of the positive-parity $\mathbf{6}_{F}$ singly heavy baryons, which actually performs a tomography for the effective strong interaction of the RQM and the spectral structure of these baryons and links the force and structure so closely.
This might be an advantage of the RQM and the improved calculation scheme over all other models and theories. The mixing effect between these orbital excited states with the same $J^{P}$ for the positive-parity $\mathbf{6}_{F}$ singly heavy baryons is found to be relatively smaller than that for the negative-party $\mathbf{6}_{F}$ singly heavy baryons, which might be the reason why the diquark approximation sometimes succeeds in the study of the singly heavy baryons. In fact, there is no mixing effect for the singly heavy baryons of the $\mathbf{\bar{3}}_{F}$ sector. Therefore, there is a big difference in the mixing effect for the different sectors of the singly heavy baryons and the flavor symmetry and parity make the difference.

In the low-lying mass spectra of singly heavy baryons, the baryon states are labeled with (mass)$J^{P}$ and $L$ is no longer regarded as a good quantum number for the $\mathbf{6}_{F}$ baryons. Most of the observed singly heavy baryons are assigned reasonably in this mass spectrum. The obtained MAD for the 74 baryons is about 6.88 MeV. Meanwhile, the nice match between the mass data and the predicted masses demonstrates the effectiveness of the improved calculation scheme.

However, there are still some problems in the study of the singly heavy baryons, such as the systematic deviation between the measured and calculated masses in the $\Xi'_{b(c)}$ and $\Sigma_{b(c)}$ families and the puzzle of the assignment for $\Lambda_{c}(2940)^{+}$, $\Lambda_{b}(6070)^{0}$ and $\Xi_{c}(3123)^{+}$. The exploration of these problems will undoubtedly help to deeply reveal more properties of the strong interaction in the baryon system.

The large amount of the theoretical calculation data obtained in this work provides reliable references for the related theoretical study of the strong interaction within the baryon system. The improved calculation scheme, especially the two-step GEM may be extended directly to the study of other baryon systems so as to explore more details of the strong interaction dominated by the non-perturbative QCD.

\begin{large}
\section*{Acknowledgements}
\end{large}

 This research was supported by the Natural Science Foundation of Guizhou Province-ZK[2024](General Project)650, the National Natural Science Foundation of China (Grant Nos. 11675265, 12575083), the Continuous Basic Scientific Research Project (Grant No. WDJC-2019-13) and the Leading Innovation Project (Grant No. LC 192209000701).

\begin{large}
\section*{Appendix}
\end{large}

\begin{table*}[htbp]
\begin{ruledtabular}\caption{Hamiltonian expectation value $\langle H\rangle$ and Contribution of each Hamiltonian term to the energy levels (in MeV) for the $1S$- and $1D$-wave states with $\langle H^{mode}\rangle\equiv\langle H_{0}+H^{conf}\rangle$ and $\langle H_{ij}\rangle\equiv\langle H\rangle-\langle(H-H_{ij})\rangle$. }
\begin{tabular}{c c c c c c c c c c c c c c c c c c c}
\label{tb1}
$(l_{\rho},l_{\lambda})nL(J^{P})_{j}$ & $\langle H^{mode}\rangle$ & $\{\langle H^{tens}_{12}\rangle$ & $\langle H^{tens}_{23}\rangle$ & $\langle H^{tens}_{31}\rangle\}$ & $\{\langle H^{cont}_{12}\rangle$  & $\langle H^{cont}_{23}\rangle$ & $\langle H^{cont}_{31}\rangle\}$ & $\{\langle H^{so(v)}_{12}\rangle$ & $\langle H^{so(v)}_{23}\rangle$ & $\langle H^{so(v)}_{31}\rangle\}$ & $\{\langle H^{so(s)}_{12}\rangle$ & $\langle H^{so(s)}_{23}\rangle$& $\langle H^{so(s)}_{31}\rangle\}$& $\langle H\rangle$  \\\hline
\multicolumn{15}{c}{$\Sigma_{b}$}\\\hline
$(0,0)1S(\frac{1}{2}^{+})_{1}$ & 5795.34  &$\{$ 0 & 0 & 0 $\}$&$\{$ 45.67  & -9.72 & -9.72 $\}$&$\{$ 0 & 0 & 0 $\}$&$\{$ 0 & 0 & 0 $\}$& 5820.75  \\
$(0,0)1S(\frac{3}{2}^{+})_{1}$ & 5795.34  &$\{$ 0 & 0 & 0 $\}$&$\{$ 44.11  & 4.57 & 4.57 $\}$&$\{$ 0 & 0 & 0 $\}$&$\{$ 0 & 0 & 0 $\}$& 5849.15  \\
$(0,2)1D(\frac{1}{2}^{+})_{1}$ & 6302.21  &$\{$ 0 & 0 & 0 $\}$&$\{$ 41.05  & 0.76 & 0.76 $\}$&$\{$ 0 & -26.95 & -26.95 $\}$&$\{$ 0 & 19.25 & 19.25 $\}$& 6329.76  \\
$(0,2)1D(\frac{3}{2}^{+})_{1}$ & 6302.21  &$\{$ 0 & -0.29 & -0.29 $\}$&$\{$ 41.00  & -0.37 & -0.37 $\}$&$\{$ 0 & -20.80 & -20.80 $\}$&$\{$ 0 & 18.40 & 18.40 $\}$& 6337.32  \\
$(0,2)1D(\frac{3}{2}^{+})_{2}$ & 6302.21  &$\{$ 0 & -0.06 & -0.06 $\}$&$\{$ 40.45  & -0.34 & -0.34 $\}$&$\{$ 0 & -10.57 & -10.57 $\}$&$\{$ 0 & 6.78 & 6.78 $\}$& 6334.23  \\
$(0,2)1D(\frac{5}{2}^{+})_{2}$ & 6302.21  &$\{$ 0 & 0.18 & 0.18 $\}$&$\{$ 40.35  & 0.22 & 0.22 $\}$&$\{$ 0 & -5.20 & -5.20 $\}$&$\{$ 0 & 5.99 & 5.99 $\}$& 6344.72  \\
$(0,2)1D(\frac{5}{2}^{+})_{3}$ & 6302.21  &$\{$ 0 & -0.25 & -0.25 $\}$&$\{$ 39.54  & -0.77 & -0.77 $\}$&$\{$ 0 & 10.88 & 10.88 $\}$&$\{$ 0 & -12.27 & -12.27 $\}$& 6337.61  \\
$(0,2)1D(\frac{7}{2}^{+})_{3}$ & 6302.21  &$\{$ 0 & 0.52 & 0.52 $\}$&$\{$ 39.41  & 0.54 & 0.54 $\}$&$\{$ 0 & 15.90 & 15.90 $\}$&$\{$ 0 & -13.19 & -13.19 $\}$& 6350.66  \\\hline
\multicolumn{15}{c}{$\Xi'_{b}$}\\\hline
$(0,0)1S(\frac{1}{2}^{+})_{1}$ & 5928.88  &$\{$ 0 & 0 & 0 $\}$&$\{$ 34.24  & -9.52 & -8.54 $\}$&$\{$ 0 & 0 & 0 $\}$&$\{$ 0 & 0 & 0 $\}$& 5944.49  \\
$(0,0)1S(\frac{3}{2}^{+})_{1}$ & 5928.88  &$\{$ 0 & 0 & 0 $\}$&$\{$ 33.08  & 4.48 & 4.03 $\}$&$\{$ 0 & 0 & 0 $\}$&$\{$ 0 & 0 & 0 $\}$& 5970.89  \\
$(0,2)1D(\frac{1}{2}^{+})_{1}$ & 6434.62  &$\{$ 0 & 0 & 0 $\}$&$\{$ 30.00  & 0.42 & 0.94 $\}$&$\{$ 0 & -20.08 & -26.46 $\}$&$\{$ 0 & 14.55 & 17.30 $\}$& 6451.54  \\
$(0,2)1D(\frac{3}{2}^{+})_{1}$ & 6434.62  &$\{$ 0 & -0.30 & -0.24 $\}$&$\{$ 29.95  & -0.20 & -0.46 $\}$&$\{$ 0 & -14.02 & -21.26 $\}$&$\{$ 0 & 13.64 & 16.42 $\}$& 6458.27  \\
$(0,2)1D(\frac{3}{2}^{+})_{2}$ & 6434.62  &$\{$ 0 & -0.06 & -0.05 $\}$&$\{$ 29.61  & -0.19 & -0.44 $\}$&$\{$ 0 & -8.34 & -10.18 $\}$&$\{$ 0 & 5.18 & 6.08 $\}$& 6456.23  \\
$(0,2)1D(\frac{5}{2}^{+})_{2}$ & 6434.62  &$\{$ 0 & 0.19 & 0.16 $\}$&$\{$ 29.53  & 0.12 & 0.28 $\}$&$\{$ 0 & -3.04 & -5.62 $\}$&$\{$ 0 & 4.35 & 5.27 $\}$& 6465.73  \\
$(0,2)1D(\frac{5}{2}^{+})_{3}$ & 6434.62  &$\{$ 0 & -0.26 & -0.22 $\}$&$\{$ 29.06  & -0.43 & -1.03 $\}$&$\{$ 0 & 6.68 & 11.68 $\}$&$\{$ 0 & -8.90 & -10.72 $\}$& 6460.84  \\
$(0,2)1D(\frac{7}{2}^{+})_{3}$ & 6434.62  &$\{$ 0 & 0.53 & 0.45 $\}$&$\{$ 28.95  & 0.30 & 0.73 $\}$&$\{$ 0 & 11.78 & 15.95 $\}$&$\{$ 0 & -9.82 & -11.60 $\}$& 6472.90  \\\hline
\multicolumn{15}{c}{$\Omega_{b}$}\\\hline
$(0,0)1S(\frac{1}{2}^{+})_{1}$ & 6033.67  &$\{$ 0 & 0 & 0 $\}$&$\{$ 27.66  & -8.89 & -8.89 $\}$&$\{$ 0 & 0 & 0 $\}$&$\{$ 0 & 0 & 0 $\}$& 6043.10  \\
$(0,0)1S(\frac{3}{2}^{+})_{1}$ & 6033.67  &$\{$ 0 & 0 & 0 $\}$&$\{$ 26.63  & 4.17 & 4.17 $\}$&$\{$ 0 & 0 & 0 $\}$&$\{$ 0 & 0 & 0 $\}$& 6069.03  \\
$(0,2)1D(\frac{1}{2}^{+})_{1}$ & 6536.09  &$\{$ 0 & 0 & 0 $\}$&$\{$ 23.48  & 0.62 & 0.62 $\}$&$\{$ 0 & -22.12 & -22.12 $\}$&$\{$ 0 & 13.66 & 13.66 $\}$& 6544.34  \\
$(0,2)1D(\frac{3}{2}^{+})_{1}$ & 6536.09  &$\{$ 0 & -0.29 & -0.29 $\}$&$\{$ 23.44  & -0.30 & -0.30 $\}$&$\{$ 0 & -16.05 & -16.05 $\}$&$\{$ 0 & 12.70 & 12.70 $\}$& 6551.96  \\
$(0,2)1D(\frac{3}{2}^{+})_{2}$ & 6536.09  &$\{$ 0 & -0.06 & -0.06 $\}$&$\{$ 23.24  & -0.28 & -0.28 $\}$&$\{$ 0 & -8.99 & -8.99 $\}$&$\{$ 0 & 4.87 & 4.87 $\}$& 6550.50  \\
$(0,2)1D(\frac{5}{2}^{+})_{2}$ & 6536.09  &$\{$ 0 & 0.18 & 0.18 $\}$&$\{$ 23.18  & 0.18 & 0.18 $\}$&$\{$ 0 & -3.70 & -3.70 $\}$&$\{$ 0 & 4.00 & 4.00 $\}$& 6560.51  \\
$(0,2)1D(\frac{5}{2}^{+})_{3}$ & 6536.09  &$\{$ 0 & -0.26 & -0.26 $\}$&$\{$ 22.91  & -0.65 & -0.65 $\}$&$\{$ 0 & 8.00 & 8.00 $\}$&$\{$ 0 & -8.16 & -8.16 $\}$& 6557.10  \\
$(0,2)1D(\frac{7}{2}^{+})_{3}$ & 6536.09  &$\{$ 0 & 0.52 & 0.52 $\}$&$\{$ 22.82  & 0.45 & 0.45 $\}$&$\{$ 0 & 13.01 & 13.01 $\}$&$\{$ 0 & -9.08 & -9.08 $\}$& 6569.54  \\\hline
\multicolumn{15}{c}{$\Sigma_{c}$}\\\hline
$(0,0)1S(\frac{1}{2}^{+})_{1}$ & 2464.30  &$\{$ 0 & 0 & 0 $\}$&$\{$ 48.04  & -27.58 & -27.58 $\}$&$\{$ 0 & 0 & 0 $\}$&$\{$ 0 & 0 & 0 $\}$& 2456.24  \\
$(0,0)1S(\frac{3}{2}^{+})_{1}$ & 2464.30  &$\{$ 0 & 0 & 0 $\}$&$\{$ 44.24  & 11.93 & 11.93 $\}$&$\{$ 0 & 0 & 0 $\}$&$\{$ 0 & 0 & 0 $\}$& 2533.92  \\
$(0,2)1D(\frac{1}{2}^{+})_{1}$ & 3041.20  &$\{$ 0 & 0 & 0 $\}$&$\{$ 39.77  & 1.72 & 1.72 $\}$&$\{$ 0 & -42.03 & -42.03 $\}$&$\{$ 0 & 23.07 & 23.07 $\}$& 3048.14  \\
$(0,2)1D(\frac{3}{2}^{+})_{1}$ & 3041.20  &$\{$ 0 & -0.73 & -0.73 $\}$&$\{$ 39.72  & -0.79 & -0.79 $\}$&$\{$ 0 & -24.37 & -24.37 $\}$&$\{$ 0 & 16.41 & 16.41 $\}$& 3062.98  \\
$(0,2)1D(\frac{3}{2}^{+})_{2}$ & 3041.20  &$\{$ 0 & -0.14 & -0.14 $\}$&$\{$ 39.28  & -0.76 & -0.76 $\}$&$\{$ 0 & -18.78 & -18.78 $\}$&$\{$ 0 & 9.95 & 9.95 $\}$& 3061.57  \\
$(0,2)1D(\frac{5}{2}^{+})_{2}$ & 3041.20  &$\{$ 0 & 0.46 & 0.46 $\}$&$\{$ 39.22  & 0.44 & 0.44 $\}$&$\{$ 0 & -3.66 & -3.66 $\}$&$\{$ 0 & 3.90 & 3.90 $\}$& 3082.51  \\
$(0,2)1D(\frac{5}{2}^{+})_{3}$ & 3041.20  &$\{$ 0 & -0.64 & -0.64 $\}$&$\{$ 38.59  & -1.65 & -1.65 $\}$&$\{$ 0 & 9.43 & 9.43 $\}$&$\{$ 0 & -8.91 & -8.91 $\}$& 3076.68  \\
$(0,2)1D(\frac{7}{2}^{+})_{3}$ & 3041.20  &$\{$ 0 & 1.29 & 1.29 $\}$&$\{$ 38.53  & 1.05 & 1.05 $\}$&$\{$ 0 & 23.17 & 23.17 $\}$&$\{$ 0 & -15.54 & -15.54 $\}$& 3101.93  \\\hline
\multicolumn{15}{c}{$\Xi'_{c}$}\\\hline
$(0,0)1S(\frac{1}{2}^{+})_{1}$ & 2603.94  &$\{$ 0 & 0 & 0 $\}$&$\{$ 35.97  & -25.65 & -24.56 $\}$&$\{$ 0 & 0 & 0 $\}$&$\{$ 0 & 0 & 0 $\}$& 2589.18  \\
$(0,0)1S(\frac{3}{2}^{+})_{1}$ & 2603.94  &$\{$ 0 & 0 & 0 $\}$&$\{$ 33.16  & 11.11 & 10.75 $\}$&$\{$ 0 & 0 & 0 $\}$&$\{$ 0 & 0 & 0 $\}$& 2660.13  \\
$(0,2)1D(\frac{1}{2}^{+})_{1}$ & 3178.86  &$\{$ 0 & 0 & 0 $\}$&$\{$ 29.09  & 0.90 & 2.19 $\}$&$\{$ 0 & -31.96 & -42.82 $\}$&$\{$ 0 & 18.27 & 20.96 $\}$& 3176.65  \\
$(0,2)1D(\frac{3}{2}^{+})_{1}$ & 3178.86  &$\{$ 0 & -0.71 & -0.64 $\}$&$\{$ 29.02  & -0.41 & -1.01 $\}$&$\{$ 0 & -15.47 & -27.12 $\}$&$\{$ 0 & 11.67 & 14.05 $\}$& 3189.00  \\
$(0,2)1D(\frac{3}{2}^{+})_{2}$ & 3178.86  &$\{$ 0 & -0.14 & -0.13 $\}$&$\{$ 28.78  & -0.39 & -0.98 $\}$&$\{$ 0 & -15.24 & -18.56 $\}$&$\{$ 0 & 8.27 & 9.25 $\}$& 3190.16  \\
$(0,2)1D(\frac{5}{2}^{+})_{2}$ & 3178.86  &$\{$ 0 & 0.44 & 0.41 $\}$&$\{$ 28.69  & 0.23 & 0.59 $\}$&$\{$ 0 & -1.08 & -5.02 $\}$&$\{$ 0 & 2.28 & 3.00 $\}$& 3208.31  \\
$(0,2)1D(\frac{5}{2}^{+})_{3}$ & 3178.86  &$\{$ 0 & -0.62 & -0.58 $\}$&$\{$ 28.38  & -0.87 & -2.23 $\}$&$\{$ 0 & 4.24 & 11.95 $\}$&$\{$ 0 & -5.58 & -7.05 $\}$& 3206.67  \\
$(0,2)1D(\frac{7}{2}^{+})_{3}$ & 3178.86  &$\{$ 0 & 1.22 & 1.18 $\}$&$\{$ 28.30  & 0.55 & 1.47 $\}$&$\{$ 0 & 17.51 & 24.32 $\}$&$\{$ 0 & -12.10 & -13.78 $\}$& 3229.15  \\\hline
\multicolumn{15}{c}{$\Omega_{c}$}\\\hline
$(0,0)1S(\frac{1}{2}^{+})_{1}$ & 2716.24  &$\{$ 0 & 0 & 0 $\}$&$\{$ 28.97  & -24.33 & -24.33 $\}$&$\{$ 0 & 0 & 0 $\}$&$\{$ 0 & 0 & 0 $\}$& 2696.35  \\
$(0,0)1S(\frac{3}{2}^{+})_{1}$ & 2716.24  &$\{$ 0 & 0 & 0 $\}$&$\{$ 26.54  & 10.53 & 10.53 $\}$&$\{$ 0 & 0 & 0 $\}$&$\{$ 0 & 0 & 0 $\}$& 2764.91  \\
$(0,2)1D(\frac{1}{2}^{+})_{1}$ & 3288.05  &$\{$ 0 & 0 & 0 $\}$&$\{$ 22.59  & 1.33 & 1.33 $\}$&$\{$ 0 & -35.46 & -35.46 $\}$&$\{$ 0 & 17.24 & 17.24 $\}$& 3278.33  \\
$(0,2)1D(\frac{3}{2}^{+})_{1}$ & 3288.05  &$\{$ 0 & -0.69 & -0.69 $\}$&$\{$ 22.54  & -0.60 & -0.60 $\}$&$\{$ 0 & -18.45 & -18.45 $\}$&$\{$ 0 & 10.23 & 10.23 $\}$& 3292.41  \\
$(0,2)1D(\frac{3}{2}^{+})_{2}$ & 3288.05  &$\{$ 0 & -0.14 & -0.14 $\}$&$\{$ 22.41  & -0.59 & -0.59 $\}$&$\{$ 0 & -16.48 & -16.48 $\}$&$\{$ 0 & 8.03 & 8.03 $\}$& 3292.67  \\
$(0,2)1D(\frac{5}{2}^{+})_{2}$ & 3288.05  &$\{$ 0 & 0.44 & 0.44 $\}$&$\{$ 22.35  & 0.34 & 0.34 $\}$&$\{$ 0 & -1.98 & -1.98 $\}$&$\{$ 0 & 1.69 & 1.69 $\}$& 3311.34  \\
$(0,2)1D(\frac{5}{2}^{+})_{3}$ & 3288.05  &$\{$ 0 & -0.62 & -0.62 $\}$&$\{$ 22.20  & -1.31 & -1.31 $\}$&$\{$ 0 & 6.07 & 6.07 $\}$&$\{$ 0 & -4.41 & -4.41 $\}$& 3309.79  \\
$(0,2)1D(\frac{7}{2}^{+})_{3}$ & 3288.05  &$\{$ 0 & 1.23 & 1.23 $\}$&$\{$ 22.14  & 0.83 & 0.83 $\}$&$\{$ 0 & 19.45 & 19.45 $\}$&$\{$ 0 & -11.24 & -11.24 $\}$& 3332.23  \\
\end{tabular}
\end{ruledtabular}
\end{table*}

\begin{table*}[htbp]
\begin{ruledtabular}\caption{Evolution of the energy levels (in MeV) for the $1S$- and $1D$-wave states with successively adding the spin-dependent terms one by one. }
\begin{tabular}{c c c c c c c c c c c c c c c c c c c}
\label{tb2}
$(l_{\rho},l_{\lambda})nL(J^{P})_{j}$ & $\langle H^{mode}\rangle$  & $\langle\cdot+ H^{tens}\rangle$ & $\langle\cdot+ H^{cont}\rangle$  & $\langle\cdot+ H^{so(v)}\rangle$ & $\langle\cdot+ H^{so(s)}\rangle$ & $\langle H\rangle$  \\\hline
\multicolumn{7}{c}{$\Sigma_{b}$}\\\hline
$(0,0)1S(\frac{1}{2}^{+})_{1}$ &5795.34 &5795.34 &5820.75 &5820.75 &5820.75 &5820.75  \\
$(0,0)1S(\frac{3}{2}^{+})_{1}$ &5795.34 &5795.34 &5849.15 &5849.15 &5849.15 &5849.15  \\
$(0,2)1D(\frac{1}{2}^{+})_{1}$ &6302.21 &6302.21 &6343.54 &6291.13 &6329.76 &6329.76  \\
$(0,2)1D(\frac{3}{2}^{+})_{1}$ &6302.21 &6301.64 &6341.13 &6300.38 &6337.32 &6337.32  \\
$(0,2)1D(\frac{3}{2}^{+})_{2}$ &6302.21 &6302.10 &6341.56 &6320.67 &6334.23 &6334.23  \\
$(0,2)1D(\frac{5}{2}^{+})_{2}$ &6302.21 &6302.58 &6343.07 &6332.72 &6344.72 &6344.72  \\
$(0,2)1D(\frac{5}{2}^{+})_{3}$ &6302.21 &6301.67 &6340.10 &6362.09 &6337.61 &6337.61 \\
$(0,2)1D(\frac{7}{2}^{+})_{3}$ &6302.21 &6303.33 &6344.62 &6376.97 &6350.66 &6350.66 \\\hline
\multicolumn{7}{c}{$\Xi'_{b}$}\\\hline
$(0,0)1S(\frac{1}{2}^{+})_{1}$ &5928.88 &5928.88 &5944.49 &5944.49 &5944.49 &5944.49  \\
$(0,0)1S(\frac{3}{2}^{+})_{1}$ &5928.88 &5928.88 &5970.89 &5970.89 &5970.89 &5970.89  \\
$(0,2)1D(\frac{1}{2}^{+})_{1}$ &6434.62 &6434.62 &6465.15 &6419.63 &6451.54 &6451.54  \\
$(0,2)1D(\frac{3}{2}^{+})_{1}$ &6434.62 &6434.09 &6462.90 &6428.15 &6458.27 &6458.27  \\
$(0,2)1D(\frac{3}{2}^{+})_{2}$ &6434.62 &6434.51 &6463.31 &6444.97 &6456.23 &6456.23  \\
$(0,2)1D(\frac{5}{2}^{+})_{2}$ &6434.62 &6434.97 &6464.73 &6456.10 &6465.73 &6465.73  \\
$(0,2)1D(\frac{5}{2}^{+})_{3}$ &6434.62 &6434.11 &6461.95 &6480.44 &6460.84 &6460.84 \\
$(0,2)1D(\frac{7}{2}^{+})_{3}$ &6434.62 &6435.68 &6466.18 &6494.29 &6472.90 &6472.90 \\\hline
\multicolumn{7}{c}{$\Omega_{b}$}\\\hline
$(0,0)1S(\frac{1}{2}^{+})_{1}$ &6033.67 &6033.67 &6043.10 &6043.10 &6043.10 &6043.10  \\
$(0,0)1S(\frac{3}{2}^{+})_{1}$ &6033.67 &6033.67 &6069.03 &6069.03 &6069.03 &6069.03 \\
$(0,2)1D(\frac{1}{2}^{+})_{1}$ &6536.09 &6536.09 &6560.21 &6517.01 &6544.34 &6544.34  \\
$(0,2)1D(\frac{3}{2}^{+})_{1}$ &6536.09 &6535.53 &6558.12 &6526.53 &6551.96 &6551.96 \\
$(0,2)1D(\frac{3}{2}^{+})_{2}$ &6536.09 &6535.98 &6558.56 &6540.76 &6550.50 &6550.50 \\
$(0,2)1D(\frac{5}{2}^{+})_{2}$ &6536.09 &6536.47 &6559.89 &6552.51 &6560.51 &6560.51 \\
$(0,2)1D(\frac{5}{2}^{+})_{3}$ &6536.09 &6535.56 &6557.27 &6573.39 &6557.10 &6557.10 \\
$(0,2)1D(\frac{7}{2}^{+})_{3}$ &6536.09 &6537.22 &6561.30 &6587.69 &6569.54 &6569.54 \\\hline
\multicolumn{7}{c}{$\Sigma_{c}$}\\\hline
$(0,0)1S(\frac{1}{2}^{+})_{1}$ &2464.30 &2464.30 &2456.24 &2456.24 &2456.24 &2456.24  \\
$(0,0)1S(\frac{3}{2}^{+})_{1}$ &2464.30 &2464.30 &2533.92 &2533.92 &2533.92 &2533.92  \\
$(0,2)1D(\frac{1}{2}^{+})_{1}$ &3041.20 &3041.20 &3082.80 &3001.91 &3048.14 &3048.14  \\
$(0,2)1D(\frac{3}{2}^{+})_{1}$ &3041.20 &3039.78 &3077.52 &3030.05 &3062.98 &3062.98  \\
$(0,2)1D(\frac{3}{2}^{+})_{2}$ &3041.20 &3040.91 &3078.63 &3041.66 &3061.57 &3061.57  \\
$(0,2)1D(\frac{5}{2}^{+})_{2}$ &3041.20 &3042.14 &3082.00 &3074.70 &3082.51 &3082.51  \\
$(0,2)1D(\frac{5}{2}^{+})_{3}$ &3041.20 &3039.85 &3075.37 &3094.43 &3076.68 &3076.68 \\
$(0,2)1D(\frac{7}{2}^{+})_{3}$ &3041.20 &3044.01 &3085.53 &3132.96 &3101.93 &3101.93 \\\hline
\multicolumn{7}{c}{$\Xi'_{c}$}\\\hline
$(0,0)1S(\frac{1}{2}^{+})_{1}$ &2603.94 &2603.94 &2589.18 &2589.18 &2589.18 &2589.18  \\
$(0,0)1S(\frac{3}{2}^{+})_{1}$ &2603.94 &2603.94 &2660.13 &2660.13 &2660.13 &2660.13  \\
$(0,2)1D(\frac{1}{2}^{+})_{1}$ &3178.86 &3178.86 &3209.86 &3137.39 &3176.65 &3176.65  \\
$(0,2)1D(\frac{3}{2}^{+})_{1}$ &3178.86 &3177.55 &3204.96 &3163.24 &3189.00 &3189.00  \\
$(0,2)1D(\frac{3}{2}^{+})_{2}$ &3178.86 &3178.60 &3205.99 &3172.63 &3190.16 &3190.16  \\
$(0,2)1D(\frac{5}{2}^{+})_{2}$ &3178.86 &3179.73 &3209.11 &3203.02 &3208.31 &3208.31  \\
$(0,2)1D(\frac{5}{2}^{+})_{3}$ &3178.86 &3177.61 &3202.96 &3219.27 &3206.67 &3206.67  \\
$(0,2)1D(\frac{7}{2}^{+})_{3}$ &3178.86 &3181.48 &3212.40 &3255.01 &3229.15 &3229.15  \\\hline
\multicolumn{7}{c}{$\Omega_{c}$}\\\hline
$(0,0)1S(\frac{1}{2}^{+})_{1}$ &2716.24 &2716.24 &2696.35 &2696.35 &2696.35 &2696.35  \\
$(0,0)1S(\frac{3}{2}^{+})_{1}$ &2716.24 &2716.24 &2764.91 &2764.91 &2764.91 &2764.91  \\
$(0,2)1D(\frac{1}{2}^{+})_{1}$ &3288.05 &3288.05 &3312.38 &3243.83 &3278.33 &3278.33  \\
$(0,2)1D(\frac{3}{2}^{+})_{1}$ &3288.05 &3286.71 &3308.03 &3271.93 &3292.41 &3292.41  \\
$(0,2)1D(\frac{3}{2}^{+})_{2}$ &3288.05 &3287.78 &3309.08 &3276.61 &3292.67 &3292.67  \\
$(0,2)1D(\frac{5}{2}^{+})_{2}$ &3288.05 &3288.94 &3311.91 &3307.96 &3311.34 &3311.34  \\
$(0,2)1D(\frac{5}{2}^{+})_{3}$ &3288.05 &3286.78 &3306.36 &3318.60 &3309.79 &3309.79  \\
$(0,2)1D(\frac{7}{2}^{+})_{3}$ &3288.05 &3290.73 &3314.98 &3354.69 &3332.23 &3332.23  \\
\end{tabular}
\end{ruledtabular}
\end{table*}

\begin{table*}[htbp]
\begin{ruledtabular}\caption{The off-diagonal matrix elements of each Hamiltonian term for the positive-parity bottom baryon states. In the $\frac{1}{2}^{+}$ subspace, the four states are abbreviated as $|(0,0)1S(\frac{1}{2}^{+})_{1}\rangle\equiv|1\rangle$, $|(0,0)2S(\frac{1}{2}^{+})_{1}\rangle\equiv|2\rangle$, $|(0,0)3S(\frac{1}{2}^{+})_{1}\rangle\equiv|3\rangle$  and $|(0,2)1D(\frac{1}{2}^{+})_{1}\rangle\equiv|4\rangle$, respectively. In the $\frac{3}{2}^{+}$ subspace, they are labeled as $|(0,0)1S(\frac{3}{2}^{+})_{1}\rangle\equiv|1'\rangle$, $|(0,0)2S(\frac{3}{2}^{+})_{1}\rangle\equiv|2'\rangle$, $|(0,0)3S(\frac{3}{2}^{+})_{1}\rangle\equiv|3'\rangle$, $|(0,2)1D(\frac{3}{2}^{+})_{1}\rangle\equiv|4'\rangle$ and $|(0,2)1D(\frac{3}{2}^{+})_{2}\rangle\equiv|5'\rangle$, respectively. And in the $\frac{5}{2}^{+}$ subspace, they are labeled as $|(0,2)1D(\frac{5}{2}^{+})_{2}\rangle\equiv|1''\rangle$ and $|(0,2)1D(\frac{5}{2}^{+})_{3}\rangle\equiv|2''\rangle$, respectively. }
\begin{tabular}{c c c c c c c c c c c c c c c c c c c c c c c c c c c c c c}
\label{tb3}
&$\langle1|\hat{O}|4\rangle$  & $\langle2|\hat{O}|4\rangle$  & $\langle3|\hat{O}|4\rangle$ & $\langle1'|\hat{O}|4'\rangle$  & $\langle2'|\hat{O}|4'\rangle$  & $\langle3'|\hat{O}|4'\rangle$ &$\langle1'|\hat{O}|5'\rangle$  & $\langle2'|\hat{O}|5'\rangle$  & $\langle3'|\hat{O}|5'\rangle$ & $\langle4'|\hat{O}|5'\rangle$ & $\langle1''|\hat{O}|2''\rangle$ \\ \hline
$\hat{O}_{ij}$ & \multicolumn{3}{c}{$\Sigma_{b}(\frac{1}{2}^{+})$} & \multicolumn{7}{c}{$\Sigma_{b}(\frac{3}{2}^{+})$} & \multicolumn{1}{c}{$\Sigma_{b}(\frac{5}{2}^{+})$} \\ \cline{1-1} \cline{2-4} \cline{5-11} \cline{12-12}
$\hat{H}_{12}^{tens}$ & 0  & 0  & 0   & 0  & 0  & 0 & 0  & 0  & 0   & 0 & 0\\
$\hat{H}_{23(31)}^{tens}$ & 1.74  & -1.22  & -0.80   & 0.89  & 0.56  & 0.41 & -2.61  & -1.69  & -1.23   & -1.94 & -1.54 \\\hline
$\hat{H}_{12}^{cont}$ & 0  & 0  & 0   & 0  & 0  & 0 & 0  & 0  & 0   & 0 & 0 \\
$\hat{H}_{23(31)}^{cont}$  & 0  & 0  & 0   & 0  & 0  & 0 & 0  & 0  & 0   & 1.06 & 0.76 \\\hline
$\hat{H}_{12}^{so(v)-2B}$  & 0  & 0  & 0   & 0  & 0  & 0 & 0  & 0  & 0   & 0 & 0   \\
$\hat{H}_{23(31)}^{so(v)-2B}$  & 0  & 0  & 0   & 0  & 0  & 0 & 0  & 0  & 0   & 1.96 & 1.50 \\\hline
$\hat{H}_{12}^{so(s)-2B}$ & 0  & 0  & 0   & 0  & 0  & 0  & 0  & 0  & 0   & 0 & 0  \\
$\hat{H}_{23(31)}^{so(s)-2B}$ & 0  & 0  & 0   & 0  & 0  & 0  & 0  & 0  & 0   & -0.29 & -0.24 \\\hline
$\hat{H}_{12}^{so(s)-3B}$ & 0 & 0 & 0 & 0 & 0 & 0 & 0 & 0 & 0 & 0 & 0   \\
$\hat{H}_{23(31)}^{so(s)-3B}$ & 0 & 0 & 0  &0 & 0 & 0  & 0 & 0 & 0 & 0.02 & 0.02 \\\hline
$\hat{H}_{12}^{so(v)-3B}$ & 0 & 0 & 0 & 0 & 0 & 0 & 0 & 0 & 0 & 0 & 0   \\
$\hat{H}_{23(31)}^{so(v)-3B}$ & 0 & 0 & 0  &0 & 0 & 0  & 0 & 0 & 0 & -0.58 & -0.43 \\\hline
Total & 3.48 & -2.45 & -1.61 & 1.78 & 1.12 & 0.83 & -5.21 & -3.38 & -2.47 & 0.48 & 0.14 \\\hline\hline
$\hat{O}_{ij}$ & \multicolumn{3}{c}{$\Xi'_{b}(\frac{1}{2}^{+})$} & \multicolumn{7}{c}{$\Xi'_{b}(\frac{3}{2}^{+})$} & \multicolumn{1}{c}{$\Xi'_{b}(\frac{5}{2}^{+})$} \\ \cline{1-1} \cline{2-4} \cline{5-11} \cline{12-12}
$\hat{H}_{12}^{tens}$ & 0  & 0  & 0   & 0  & 0  & 0 & 0  & 0  & 0   & 0 & 0 \\
$\hat{H}_{23}^{tens}$ & -0.78  & 0.63  & 0.83   & -0.40  & 0.28  & -0.41 & 1.19  & -0.82  & 1.23   & -1.22 & 0.77 \\
$\hat{H}_{31}^{tens}$ &-2.41 & 2.08 & 0.41   & -1.23  & 0.98     & -0.24 & 3.64 & -2.98 & 0.71 & -2.89 & 2.64 \\\hline
$\hat{H}_{12}^{cont}$ & 0  & 0  & 0   & 0  & 0  & 0 & 0  & 0  & 0   & 0 & 0 \\
$\hat{H}_{23}^{cont}$& 0  & 0  & 0   & 0  & 0  & 0 & 0  & 0  & 0   & 0.58 & -0.42 \\
$\hat{H}_{31}^{cont}$ & 0  & 0  & 0   & 0  & 0  & 0 & 0  & 0  & 0   & 1.34 & -1.00 \\\hline
$\hat{H}_{12}^{so(v)-2B}$ & 0  & 0  & 0   & 0  & 0  & 0 & 0  & 0  & 0   & 0 & 0 \\
$\hat{H}_{23}^{so(v)-2B}$ & 0  & 0  & 0   & 0  & 0  & 0 & 0  & 0  & 0   & 1.95 & -1.49 \\
$\hat{H}_{31}^{so(v)-2B}$ & 0  & 0  & 0   & 0  & 0  & 0 & 0  & 0  & 0   & 1.65 & -1.28 \\\hline
$\hat{H}_{12}^{so(s)-2B}$ & 0  & 0  & 0   & 0  & 0  & 0 & 0  & 0  & 0   & 0 & 0 \\
$\hat{H}_{23}^{so(s)-2B}$ & 0  & 0  & 0   & 0  & 0  & 0 & 0  & 0  & 0   & -0.31 & 0.25 \\
$\hat{H}_{31}^{so(s)-2B}$ & 0  & 0  & 0   & 0  & 0  & 0 & 0  & 0  & 0   & -0.29 & 0.24 \\\hline
$\hat{H}_{12}^{so(s)-3B}$ & 0  & 0  & 0   & 0  & 0  & 0 & 0  & 0  & 0   & 0 & 0 \\
$\hat{H}_{23}^{so(s)-3B}$ & 0  & 0  & 0   & 0  & 0  & 0 & 0  & 0  & 0   & 0.01 & -0.01 \\
$\hat{H}_{31}^{so(s)-3B}$ & 0  & 0  & 0   & 0  & 0  & 0 & 0  & 0  & 0   & 0.03 & -0.03 \\\hline
$\hat{H}_{12}^{so(v)-3B}$ & 0  & 0  & 0   & 0  & 0  & 0 & 0  & 0  & 0   & 0 & 0 \\
$\hat{H}_{23}^{so(v)-3B}$ & 0  & 0  & 0   & 0  & 0  & 0 & 0  & 0  & 0   & -0.34 & 0.25 \\
$\hat{H}_{31}^{so(v)-3B}$ & 0  & 0  & 0   & 0  & 0  & 0 & 0  & 0  & 0   & -0.47 & 0.36 \\ \hline
Total & -3.19 & 2.71 & 1.24 & -1.64 & 1.26 & -0.65  & 4.83 & -3.81 & 1.94 & 0.05 &0.27 \\ \hline \hline
$\hat{O}_{ij}$ & \multicolumn{3}{c}{$\Omega_{b}(\frac{1}{2}^{+})$} & \multicolumn{7}{c}{$\Omega_{b}(\frac{3}{2}^{+})$} & \multicolumn{1}{c}{$\Omega_{b}(\frac{5}{2}^{+})$} \\ \cline{1-1} \cline{2-4} \cline{5-11} \cline{12-12}
$\hat{H}_{12}^{tens}$ & 0  & 0  & 0   & 0  & 0  & 0 & 0  & 0  & 0   & 0 & 0\\
$\hat{H}_{23(31)}^{tens}$ & 1.81  & 1.35  & -1.06   & 0.92  & 0.61  & 0.54 & 2.72  & 1.82  & 1.59   & 2.11 & -1.69 \\\hline
$\hat{H}_{12}^{cont}$ & 0  & 0  & 0   & 0  & 0  & 0 & 0  & 0  & 0   & 0 & 0 \\
$\hat{H}_{23(31)}^{cont}$  & 0  & 0  & 0   & 0  & 0  & 0 & 0  & 0  & 0   & -0.86 & 0.64 \\\hline
$\hat{H}_{12}^{so(v)-2B}$  & 0  & 0  & 0   & 0  & 0  & 0 & 0  & 0  & 0   & 0 & 0   \\
$\hat{H}_{23(31)}^{so(v)-2B}$  & 0  & 0  & 0   & 0  & 0  & 0 & 0  & 0  & 0   & -1.93 & 1.48 \\\hline
$\hat{H}_{12}^{so(s)-2B}$ & 0  & 0  & 0   & 0  & 0  & 0  & 0  & 0  & 0   & 0 & 0  \\
$\hat{H}_{23(31)}^{so(s)-2B}$ & 0  & 0  & 0   & 0  & 0  & 0  & 0  & 0  & 0   & 0.32 & -0.25 \\\hline
$\hat{H}_{12}^{so(s)-3B}$ & 0 & 0 & 0 & 0 & 0 & 0 & 0 & 0 & 0 & 0 & 0   \\
$\hat{H}_{23(31)}^{so(s)-3B}$ & 0 & 0 & 0  &0 & 0 & 0  & 0 & 0 & 0 & -0.03 & 0.02 \\\hline
$\hat{H}_{12}^{so(v)-3B}$ & 0 & 0 & 0 & 0 & 0 & 0 & 0 & 0 & 0 & 0 & 0   \\
$\hat{H}_{23(31)}^{so(v)-3B}$ & 0 & 0 & 0  &0 & 0 & 0  & 0 & 0 & 0 & 0.58 & -0.44 \\\hline
Total & 3.62 & 2.69 & -2.12 & 1.84 & 1.22 & 1.08 & 5.43 & 3.64 & 3.19 & 0.36 & -0.49 \\
\end{tabular}
\end{ruledtabular}
\end{table*}

\begin{table*}[htbp]
\begin{ruledtabular}\caption{Same as Table~\ref{tb3}, but for the charm baryon states. }
\begin{tabular}{c c c c c c c c c c c c c c c c c c c c c c c c c c c c c c}
\label{tb4}
&$\langle1|\hat{O}|4\rangle$  & $\langle2|\hat{O}|4\rangle$  & $\langle3|\hat{O}|4\rangle$ & $\langle1'|\hat{O}|4'\rangle$  & $\langle2'|\hat{O}|4'\rangle$  & $\langle3'|\hat{O}|4'\rangle$ &$\langle1'|\hat{O}|5'\rangle$  & $\langle2'|\hat{O}|5'\rangle$  & $\langle3'|\hat{O}|5'\rangle$ & $\langle4'|\hat{O}|5'\rangle$ & $\langle1''|\hat{O}|2''\rangle$ \\ \hline
$\hat{O}$ & \multicolumn{3}{c}{$\Sigma_{c}(\frac{1}{2}^{+})$} & \multicolumn{7}{c}{$\Sigma_{c}(\frac{3}{2}^{+})$} & \multicolumn{1}{c}{$\Sigma_{c}(\frac{5}{2}^{+})$} \\ \cline{1-1} \cline{2-4} \cline{5-11} \cline{12-12}
$\hat{H}_{12}^{tens}$ & 0  & 0  & 0   & 0  & 0  & 0 & 0  & 0  & 0   & 0 & 0\\
$\hat{H}_{23(31)}^{tens}$ & 4.53  & 3.65  & 2.08   & 2.35  & -1.50  & -1.15 & 6.97  & -4.54  & -3.45   & 4.72 & -3.60 \\ \hline
$\hat{H}_{12}^{cont}$ & 0  & 0  & 0   & 0  & 0  & 0 & 0  & 0  & 0   & 0 & 0 \\
$\hat{H}_{23(31)}^{cont}$  & 0  & 0  & 0   & 0  & 0  & 0 & 0  & 0  & 0   & -2.32 & 1.61 \\ \hline
$\hat{H}_{12}^{so(v)-2B}$  & 0  & 0  & 0   & 0  & 0  & 0 & 0  & 0  & 0   & 0 & 0   \\
$\hat{H}_{23(31)}^{so(v)-2B}$  & 0  & 0  & 0   & 0  & 0  & 0 & 0  & 0  & 0   & -5.59 & 4.17 \\ \hline
$\hat{H}_{12}^{so(s)-2B}$ & 0  & 0  & 0   & 0  & 0  & 0  & 0  & 0  & 0   & 0 & 0  \\
$\hat{H}_{23(31)}^{so(s)-2B}$ & 0  & 0  & 0   & 0  & 0  & 0  & 0  & 0  & 0   & 2.23 & -1.78 \\ \hline
$\hat{H}_{12}^{so(s)-3B}$ & 0 & 0 & 0 & 0 & 0 & 0 & 0 & 0 & 0 & 0 & 0   \\
$\hat{H}_{23(31)}^{so(s)-3B}$ & 0 & 0 & 0  &0 & 0 & 0  & 0 & 0 & 0 & -0.17 & 0.13 \\ \hline
$\hat{H}_{12}^{so(v)-3B}$ & 0 & 0 & 0 & 0 & 0 & 0 & 0 & 0 & 0 & 0 & 0   \\
$\hat{H}_{23(31)}^{so(v)-3B}$ & 0 & 0 & 0  &0 & 0 & 0  & 0 & 0 & 0 & 1.57 & -1.14 \\ \hline
Total & 9.07 & 7.29 & 4.17 & 4.71 & -3.00 & -2.31 & 13.95 & -9.07 & -6.91 & 0.89 & -1.24 \\ \hline \hline
$\hat{O}$ & \multicolumn{3}{c}{$\Xi'_{c}(\frac{1}{2}^{+})$} & \multicolumn{7}{c}{$\Xi'_{c}(\frac{3}{2}^{+})$} & \multicolumn{1}{c}{$\Xi'_{c}(\frac{5}{2}^{+})$} \\ \cline{1-1} \cline{2-4} \cline{5-11} \cline{12-12}
$\hat{H}_{12}^{tens}$ & 0  & 0  & 0   & 0  & 0  & 0 & 0  & 0  & 0   & 0 & 0 \\
$\hat{H}_{23}^{tens}$ & -1.88  & -1.85  & -1.88   & -0.99  & 0.73  & 0.97 & 2.95  & -2.18  & -2.90   & -2.72 & -1.62 \\
$\hat{H}_{31}^{tens}$ & -6.55 & -5.89 & -1.01   & -3.44  & 2.59 & 0.74 & 10.22 & -7.82 & -2.21 & -7.30 & -6.45 \\ \hline
$\hat{H}_{12}^{cont}$ & 0  & 0  & 0   & 0  & 0  & 0 & 0  & 0  & 0   & 0 & 0 \\
$\hat{H}_{23}^{cont}$& 0  & 0  & 0   & 0  & 0  & 0 & 0  & 0  & 0   & 1.20 & 0.84 \\
$\hat{H}_{31}^{cont}$ & 0  & 0  & 0   & 0  & 0  & 0 & 0  & 0  & 0   & 3.00 & 2.15 \\ \hline
$\hat{H}_{12}^{so(v)-2B}$ & 0  & 0  & 0   & 0  & 0  & 0 & 0  & 0  & 0   & 0 & 0 \\
$\hat{H}_{23}^{so(v)-2B}$ & 0  & 0  & 0   & 0  & 0  & 0 & 0  & 0  & 0   & 5.26 & 3.94 \\
$\hat{H}_{31}^{so(v)-2B}$ & 0  & 0  & 0   & 0  & 0  & 0 & 0  & 0  & 0   & 4.95 & 3.75 \\ \hline
$\hat{H}_{12}^{so(s)-2B}$ & 0  & 0  & 0   & 0  & 0  & 0 & 0  & 0  & 0   & 0 & 0 \\
$\hat{H}_{23}^{so(s)-2B}$ & 0  & 0  & 0   & 0  & 0  & 0 & 0  & 0  & 0   & -2.19 & -1.76 \\
$\hat{H}_{31}^{so(s)-2B}$ & 0  & 0  & 0   & 0  & 0  & 0 & 0  & 0  & 0   & -2.28 & -1.83 \\ \hline
$\hat{H}_{12}^{so(s)-3B}$ & 0  & 0  & 0   & 0  & 0  & 0 & 0  & 0  & 0   & 0 & 0 \\
$\hat{H}_{23}^{so(s)-3B}$ & 0  & 0  & 0   & 0  & 0  & 0 & 0  & 0  & 0   & 0.09 & 0.06 \\
$\hat{H}_{31}^{so(s)-3B}$ & 0  & 0  & 0   & 0  & 0  & 0 & 0  & 0  & 0   & 0.25 & 0.19 \\ \hline
$\hat{H}_{12}^{so(v)-3B}$ & 0  & 0  & 0   & 0  & 0  & 0 & 0  & 0  & 0   & 0 & 0 \\
$\hat{H}_{23}^{so(v)-3B}$ & 0  & 0  & 0   & 0  & 0  & 0 & 0  & 0  & 0   & -0.84 & -0.60 \\
$\hat{H}_{31}^{so(v)-3B}$ & 0  & 0  & 0   & 0  & 0  & 0 & 0  & 0  & 0   & -1.46 & -1.07 \\ \hline
Total & -8.43 & -7.74 & -2.89 & -4.43 & 3.32 & 1.71  & 13.17 & -10.00 & -5.10 & -2.04 &-2.42 \\ \hline \hline
$\hat{O}$ & \multicolumn{3}{c}{$\Omega_{c}(\frac{1}{2}^{+})$} & \multicolumn{7}{c}{$\Omega_{c}(\frac{3}{2}^{+})$} & \multicolumn{1}{c}{$\Omega_{c}(\frac{5}{2}^{+})$} \\ \cline{1-1} \cline{2-4} \cline{5-11} \cline{12-12}
$\hat{H}_{12}^{tens}$ & 0  & 0  & 0   & 0  & 0  & 0 & 0  & 0  & 0   & 0 & 0\\
$\hat{H}_{23(31)}^{tens}$ & -4.45  & 3.75& 2.59&-2.29&1.53  &-1.37 &-6.82  & 4.61  & -4.10 & 4.70 & -3.58 \\\hline
$\hat{H}_{12}^{cont}$ & 0  & 0  & 0   & 0  & 0  & 0 & 0  & 0  & 0   & 0 & 0 \\
$\hat{H}_{23(31)}^{cont}$  & 0  & 0  & 0   & 0  & 0  & 0 & 0  & 0  & 0   & -1.78 & 1.26 \\\hline
$\hat{H}_{12}^{so(v)-2B}$  & 0  & 0  & 0   & 0  & 0  & 0 & 0  & 0  & 0   & 0 & 0   \\
$\hat{H}_{23(31)}^{so(v)-2B}$  & 0  & 0  & 0   & 0  & 0  & 0 & 0  & 0  & 0   & -5.35 & 4.01 \\\hline
$\hat{H}_{12}^{so(s)-2B}$ & 0  & 0  & 0   & 0  & 0  & 0  & 0  & 0  & 0   & 0 & 0  \\
$\hat{H}_{23(31)}^{so(s)-2B}$ & 0  & 0  & 0   & 0  & 0  & 0  & 0  & 0  & 0   & 0.32 & -0.25 \\\hline
$\hat{H}_{12}^{so(s)-3B}$ & 0 & 0 & 0 & 0 & 0 & 0 & 0 & 0 & 0 & 0 & 0   \\
$\hat{H}_{23(31)}^{so(s)-3B}$ & 0 & 0 & 0  &0 & 0 & 0  & 0 & 0 & 0 & -0.17 & 0.13 \\\hline
$\hat{H}_{12}^{so(v)-3B}$ & 0 & 0 & 0 & 0 & 0 & 0 & 0 & 0 & 0 & 0 & 0   \\
$\hat{H}_{23(31)}^{so(v)-3B}$ & 0 & 0 & 0  &0 & 0 & 0  & 0 & 0 & 0 & 1.50 & -1.09 \\\hline
Total & -8.90 & 7.51 & 5.17 & -4.58 & 3.07 & -2.74 & -13.64 & 9.21 & -8.20 & 2.43 & -2.27 \\
\end{tabular}
\end{ruledtabular}
\end{table*}

\begin{table*}[htbp]
\begin{ruledtabular}\caption{The Hamiltonian matrix $H$ (in MeV), eigenvalues (in MeV) and eigenvectors in the $\frac{1}{2}^{+}$, $\frac{3}{2}^{+}$, $\frac{5}{2}^{+}$ and $\frac{7}{2}^{+}$ subspaces of the bottom baryon states. The elements in boldface in the eigenvectors correspond to the main contributors of the orbitally excited states, namely the special orbitally excited states. }
\begin{tabular}{c c c c c c c c c c c c c c c c c c c}
\label{tb5}
$(J^{P})$ &$(l_{\rho},l_{\lambda})nL(J^{P})_{j}$ & $H$ & Eigenvalue & Eigenvector   \\\hline
\multirow{4}{*}{$\Sigma_{b}(\frac{1}{2}^{+})$} &$(0,0)1S(\frac{1}{2}^{+})_{1}$ & \multirow{4}{*}{$ \begin{pmatrix} 5820.75 & 0 & 0 & 3.48 \\ 0 & 6226.18 & 0 & -2.45 \\ 0 & 0 & 6438.63 & -1.61\\ 3.48&-2.45 & -1.61 & 6329.76 \end{pmatrix} $}  &5820.73 & (\textbf{-0.99998}, 0.00004, 0.00002, 0.00684)\\
     & $(0,0)2S(\frac{1}{2}^{+})_{1}$ & &6226.12 &(-0.00020, -\textbf{0.99972}, -0.00018, -0.02363)\\
     & $(0,0)3S(\frac{1}{2}^{+})_{1}$ & &6329.82 &(0.00683, -0.02362, 0.01479, \textbf{0.99959})\\
    & $(0,2)1D(\frac{1}{2}^{+})_{1}$ & &6438.65 &(0.00008, -0.00017, \textbf{0.99989}, -0.01479)\\\hline
\multirow{5}{*}{$\Sigma_{b}(\frac{3}{2}^{+})$} &$(0,0)1S(\frac{3}{2}^{+})_{1}$ & \multirow{5}{*}{$ \begin{pmatrix} 5849.15 & 0& 0 &1.78&-5.21\\ 0& 6245.90&0& 1.12& -3.38 \\ 0& 0& 6446.14& 0.83&-2.47 \\1.78& 1.12&0.83&6337.32&0.48 \\ -5.21&-3.38&-2.47&0.48&6334.23\end{pmatrix} $}  &5849.09 & (\textbf{0.99994}, 0.00010, 0.00005, -0.00366, 0.01074)\\
     & $(0,0)2S(\frac{3}{2}^{+})_{1}$ & &6245.76 &(0.00056, -\textbf{0.99919}, -0.00052, 0.01242, -0.03822)\\
      & $(0,0)3S(\frac{3}{2}^{+})_{1}$ & &6334.30 &(-0.01114, -0.03961, 0.02289, -0.14209, \textbf{0.98873})\\
       & $(0,2)1D(\frac{3}{2}^{+})_{1}$ & &6337.40 &(0.00209, 0.00685, -0.00432, \textbf{0.98974}, 0.14263)\\
    & $(0,2)1D(\frac{3}{2}^{+})_{2}$ & &6446.20 &(0.00022, 0.00041, \textbf{0.99973}, 0.00753, -0.02204)\\\hline
\multirow{2}{*}{$\Sigma_{b}(\frac{5}{2}^{+})$} &$(0,2)1D(\frac{5}{2}^{+})_{2}$ & \multirow{2}{*}{$ \begin{pmatrix} 6344.72 & 0.14 \\ 0.14& 6337.61\end{pmatrix} $}  &6337.61 & (0.019679, -\textbf{0.999806})\\
     & $(0,2)1D(\frac{5}{2}^{+})_{3}$ & &6344.72 &(-\textbf{0.999806}, -0.019679)\\\hline
$\Sigma_{b}(\frac{7}{2}^{+})$ & $(0,2)1D(\frac{7}{2}^{+})_{3}$ & (6350.66) & 6350.66 & (1) \\ \hline \hline
\multirow{4}{*}{$\Xi'_{b}(\frac{1}{2}^{+})$} &$(0,0)1S(\frac{1}{2}^{+})_{1}$ & \multirow{4}{*}{$ \begin{pmatrix} 5944.49 & 0 & 0 & -3.19 \\ 0 & 6351.02 & 0 & 2.71 \\ 0 & 0 & 6542.97 & 1.24\\ -3.19&2.71 & 1.24 & 6451.54 \end{pmatrix} $}  &5944.47 & (-\textbf{0.999980}, 0.000042, 0.000013, -0.006291)\\
     & $(0,0)2S(\frac{1}{2}^{+})_{1}$ & &6350.95 &(-0.000211, -\textbf{0.999637}, -0.000174, 0.026926)\\
     & $(0,0)3S(\frac{1}{2}^{+})_{1}$ & &6451.62 &(-0.006287, 0.026927, -0.013567, \textbf{0.999526})\\
    & $(0,2)1D(\frac{1}{2}^{+})_{1}$ & &6542.99 &(-0.000072, 0.000192, \textbf{0.999908}, 0.013567)\\\hline
\multirow{5}{*}{$\Xi'_{b}(\frac{3}{2}^{+})$} &$(0,0)1S(\frac{3}{2}^{+})_{1}$ & \multirow{5}{*}{$ \begin{pmatrix} 5970.89 & 0& 0 &-1.64&4.83\\ 0& 6369.31&0& 1.26& -3.81 \\ 0& 0& 6550.94& -0.65&1.94 \\-1.64& 1.26&-0.65 & 6458.27 &0.05 \\ 4.83 &-3.81 &1.94 &0.05 &6456.23 \end{pmatrix} $}  &5970.84 & (-\textbf{0.99995}, 0.00011, -0.00004, -0.00337, 0.00995)\\
     & $(0,0)2S(\frac{3}{2}^{+})_{1}$ & &6369.13 &(0.00059, \textbf{0.99895}, -0.00052, -0.01414, 0.04368)\\
      & $(0,0)3S(\frac{3}{2}^{+})_{1}$ & &6456.40 &(0.00992, -0.04363, -0.02047, 0.00429, \textbf{0.99878})\\
       & $(0,2)1D(\frac{3}{2}^{+})_{1}$ & &6458.29 &(0.00340, -0.01431, -0.00709, -\textbf{0.99986}, 0.00349)\\
    & $(0,2)1D(\frac{3}{2}^{+})_{2}$ & &6550.98 &(-0.00019, 0.00048, -\textbf{0.99977}, 0.00701, -0.02049)\\\hline
\multirow{2}{*}{$\Xi'_{b}(\frac{5}{2}^{+})$} &$(0,2)1D(\frac{5}{2}^{+})_{2}$ & \multirow{2}{*}{$ \begin{pmatrix} 6465.73 & 0.27 \\ 0.27& 6460.84\end{pmatrix} $}  &6460.83 & (0.0549642, -\textbf{0.998488})\\
     & $(0,2)1D(\frac{5}{2}^{+})_{3}$ & &6465.74 &(-\textbf{0.998488}, -0.0549642)\\\hline
$\Xi'_{b}(\frac{7}{2}^{+})$ & $(0,2)1D(\frac{7}{2}^{+})_{3}$ & (6472.90) & 6472.90 & (1) \\ \hline \hline
  \multirow{4}{*}{$\Omega_{b}(\frac{1}{2}^{+})$} &$(0,0)1S(\frac{1}{2}^{+})_{1}$ & \multirow{4}{*}{$ \begin{pmatrix} 6043.10 & 0 & 0 & 3.62 \\ 0 & 6447.79 & 0 & 2.69 \\ 0 & 0 & 6640.56 & -2.12\\ 3.62 &2.69 & -2.12 & 6544.34 \end{pmatrix} $}  &6043.07 & (\textbf{0.999974}, 0.000048, -0.000026, -0.007222)\\
     & $(0,0)2S(\frac{1}{2}^{+})_{1}$ & &6447.72 &(-0.000249, \textbf{0.999613}, -0.000306, -0.027826)\\
     & $(0,0)3S(\frac{1}{2}^{+})_{1}$ & &6544.39 &(0.007217, 0.027827, 0.022031, \textbf{0.999344})\\
    & $(0,2)1D(\frac{1}{2}^{+})_{1}$ & &6640.61 &(-0.000133, -0.000307, \textbf{0.999757}, -0.022030)\\\hline
    \multirow{5}{*}{$\Omega_{b}(\frac{3}{2}^{+})$} &$(0,0)1S(\frac{3}{2}^{+})_{1}$ & \multirow{5}{*}{$ \begin{pmatrix} 6069.03& 0& 0 &1.84 &5.43\\ 0& 6465.47 &0& 1.22 & 3.64 \\ 0& 0& 6647.24 &1.08 &3.19 \\1.84& 1.22 &1.08 & 6551.96 &0.36 \\ 5.43 &3.64 &3.19 &0.36 &6550.50 \end{pmatrix} $}  &6068.96 & (\textbf{0.99993}, 0.00012, 0.00007, -0.00380, -0.01127)\\
     & $(0,0)2S(\frac{3}{2}^{+})_{1}$ & &6465.3 &(0.00065, -\textbf{0.99900}, -0.00083, 0.01388, 0.04261)\\
      & $(0,0)3S(\frac{3}{2}^{+})_{1}$ & &6550.5 &(0.00988, 0.03753, -0.02888, -0.26095, \textbf{0.96414})\\
       & $(0,2)1D(\frac{3}{2}^{+})_{1}$ & &6552.08 &(-0.00659, -0.02451, 0.01966, -\textbf{0.96518}, -0.25962)\\
    & $(0,2)1D(\frac{3}{2}^{+})_{2}$ & &6647.36 &(0.00035, 0.00074, \textbf{0.99939}, 0.01145, 0.03300)\\\hline
    \multirow{2}{*}{$\Omega_{b}(\frac{5}{2}^{+})$} &$(0,2)1D(\frac{5}{2}^{+})_{2}$ & \multirow{2}{*}{$ \begin{pmatrix} 6560.51 & -0.49 \\ -0.49& 6557.10\end{pmatrix} $}  &6557.03 & (-0.139468, -\textbf{0.990227})\\
     & $(0,2)1D(\frac{5}{2}^{+})_{3}$ & &6560.58 &(-\textbf{0.990227}, 0.139468)\\\hline
    $\Omega_{b}(\frac{7}{2}^{+})$ & $(0,2)1D(\frac{7}{2}^{+})_{3}$ & (6569.54) & 6569.54 & (1) \\
\end{tabular}
\end{ruledtabular}
\end{table*}

\begin{table*}[htbp]
\begin{ruledtabular}\caption{Same as Table~\ref{tb5}, but for the charm baryon states. }
\begin{tabular}{c c c c c c c c c c c c c c c c c c c}
\label{tb6}
$(J^{P})$ &$(l_{\rho},l_{\lambda})nL(J^{P})_{j}$ & $H$ & Eigenvalue & Eigenvector   \\\hline
\multirow{4}{*}{$\Sigma_{c}(\frac{1}{2}^{+})$} &$(0,0)1S(\frac{1}{2}^{+})_{1}$ & \multirow{4}{*}{$ \begin{pmatrix} 2456.24 & 0& 0 & 9.07 \\ 0 & 2913.44 & 0 & 7.29 \\ 0 &0 & 3109.47 & 4.17\\9.07& 7.29 &4.17&3048.14  \end{pmatrix} $}  &2456.10 & (\textbf{-0.999883}, -0.000244, -0.000098, 0.015322)\\
     & $(0,0)2S(\frac{1}{2}^{+})_{1}$ & &2913.05 &(-0.001069, \textbf{0.998548}, 0.001143, -0.053848)\\
     & $(0,0)3S(\frac{1}{2}^{+})_{1}$ & &3048.39 &(0.015258, 0.053811, -0.068004, \textbf{0.996116})\\
    & $(0,2)1D(\frac{1}{2}^{+})_{1}$ & &3109.75 &(-0.000943, -0.002524, \textbf{-0.997684}, -0.067960)\\\hline
    \multirow{5}{*}{$\Sigma_{c}(\frac{3}{2}^{+})$} &$(0,0)1S(\frac{3}{2}^{+})_{1}$ & \multirow{5}{*}{$ \begin{pmatrix} 2533.92 & 0 & 0 & 4.71& 13.95\\ 0& 2967.28 & 0 & -3.00& -9.07\\ 0& 0& 3126.65 & -2.31&-6.91 \\ 4.71&-3.00&-2.31&3062.98 &0.89 \\ 13.95& -9.07&-6.91 &0.89&3061.57\end{pmatrix} $}  &2533.51 & (-\textbf{0.99961}, 0.00061, 0.00034, 0.00885, 0.02641)\\
     & $(0,0)2S(\frac{3}{2}^{+})_{1}$ & &2966.33 &(-0.00337, -\textbf{0.99508}, -0.00450, -0.02996, -0.09432)\\
      & $(0,0)3S(\frac{3}{2}^{+})_{1}$ & &3061.40 &(0.01729, -0.06335, 0.06941, -0.54024, \textbf{0.83607})\\
       & $(0,2)1D(\frac{3}{2}^{+})_{1}$ & &3063.69 &(-0.02141, 0.07594, -0.08891, -\textbf{0.84009}, -0.52926)\\
    & $(0,2)1D(\frac{3}{2}^{+})_{2}$ & &3127.48 &(-0.00279, 0.00672, \textbf{0.99361}, -0.03757, -0.10620)\\\hline
    \multirow{2}{*}{$\Sigma_{c}(\frac{5}{2}^{+})$} &$(0,2)1D(\frac{5}{2}^{+})_{2}$ & \multirow{2}{*}{$ \begin{pmatrix} 3082.51 & -1.24 \\ -1.24 & 3076.68\end{pmatrix} $}  &3076.43 & (-0.199746, -\textbf{0.979848})\\
     & $(0,2)1D(\frac{5}{2}^{+})_{3}$ & &3082.76 &(-\textbf{0.979848}, 0.199746)\\ \hline
    $\Sigma_{c}(\frac{7}{2}^{+})$ & $(0,2)1D(\frac{7}{2}^{+})_{3}$ & (3101.93) & 3101.93 & (1) \\ \hline \hline
    \multirow{4}{*}{$\Xi'_{c}(\frac{1}{2}^{+})$} &$(0,0)1S(\frac{1}{2}^{+})_{1}$ & \multirow{4}{*}{$ \begin{pmatrix} 2589.18 & 0& 0 & -8.43 \\ 0 & 3046.19 & 0 & -7.74 \\ 0 &0 & 3219.50 & -2.89\\-8.43 & -7.74 &-2.89 &3176.65 \end{pmatrix} $}  &2589.06 & (\textbf{0.999897}, 0.000243, 0.000066, 0.014349)\\
     & $(0,0)2S(\frac{1}{2}^{+})_{1}$ & &3045.73 &(0.001089, -\textbf{0.998259}, -0.000981, -0.058970)\\
     & $(0,0)3S(\frac{1}{2}^{+})_{1}$ & &3177.03 &(-0.014281, -0.058910, 0.067770, \textbf{0.995858})\\
    & $(0,2)1D(\frac{1}{2}^{+})_{1}$ & &3219.70 &(-0.000905, -0.003020, -\textbf{0.997701}, 0.067703)\\\hline
    \multirow{5}{*}{$\Xi'_{c}(\frac{3}{2}^{+})$} &$(0,0)1S(\frac{3}{2}^{+})_{1}$ & \multirow{5}{*}{$ \begin{pmatrix} 2660.13 & 0 & 0 & -4.43 & 13.17\\ 0& 3095.71 & 0 & -3.32 & -10.00 \\ 0& 0& 3237.15 & 1.71 &-5.10 \\ -4.43 &3.32 &1.71 &3189.00 &-2.04 \\ 13.17& -10.00 &-5.10 &-2.04 &3190.16 \end{pmatrix} $}  &2659.77 & (\textbf{0.99966}, -0.00063, -0.00024, 0.00828, -0.02480)\\
     & $(0,0)2S(\frac{3}{2}^{+})_{1}$ & &3094.56 &(-0.00346, -\textbf{0.99413}, -0.00408, 0.03264, -0.10304)\\
      & $(0,0)3S(\frac{3}{2}^{+})_{1}$ & &3187.54 &(0.00684, -0.03021, 0.02824, \textbf{0.83173}, 0.55359)\\
       & $(0,2)1D(\frac{3}{2}^{+})_{1}$ & &3192.49 &(0.02485, -0.10354, 0.11463, -0.55269, \textbf{0.81857})\\
    & $(0,2)1D(\frac{3}{2}^{+})_{2}$ & &3237.79 &(-0.00283, 0.00873, \textbf{0.99300}, 0.04028, -0.11067)\\ \hline
    \multirow{2}{*}{$\Xi'_{c}(\frac{5}{2}^{+})$} &$(0,2)1D(\frac{5}{2}^{+})_{2}$ & \multirow{2}{*}{$ \begin{pmatrix} 3208.31 & -2.42 \\ -2.42 & 3206.67\end{pmatrix} $}  &3204.93 & (-0.582700, -\textbf{0.812687})\\
     & $(0,2)1D(\frac{5}{2}^{+})_{3}$ & &3210.05 &(-\textbf{0.812687}, 0.582700)\\\hline
    $\Xi'_{c}(\frac{7}{2}^{+})$ & $(0,2)1D(\frac{7}{2}^{+})_{3}$ & (3229.15) & 3229.15 & (1) \\ \hline \hline
\multirow{4}{*}{$\Omega_{c}(\frac{1}{2}^{+})$} &$(0,0)1S(\frac{1}{2}^{+})_{1}$ & \multirow{4}{*}{$ \begin{pmatrix} 2696.35& 0& 0 & -8.90 \\ 0 & 3150.05 & 0 & 7.51 \\ 0 &0 & 3324.71 & 5.17\\-8.90 & 7.51 &5.17 &3278.33 \end{pmatrix} $}  &2696.21 & (-\textbf{0.999883}, 0.000253, 0.000126, -0.015292)\\
     & $(0,0)2S(\frac{1}{2}^{+})_{1}$ & &3149.61 &(-0.001143, -\textbf{0.998301}, -0.001719, 0.058235)\\
     & $(0,0)3S(\frac{1}{2}^{+})_{1}$ & &3278.33 &(-0.015171, 0.058079, -0.110583, \textbf{0.992052})\\
    & $(0,2)1D(\frac{1}{2}^{+})_{1}$ & &3325.28 &(-0.001563, 0.004735, \textbf{0.993865}, 0.110484)\\\hline
    \multirow{5}{*}{$\Omega_{c}(\frac{3}{2}^{+})$} &$(0,0)1S(\frac{3}{2}^{+})_{1}$ & \multirow{5}{*}{$ \begin{pmatrix} 2764.91 & 0 & 0 & -4.58 & -13.64\\ 0& 3197.69 & 0 & 3.07 & 9.21 \\ 0& 0& 3339.30 & -2.74 &-8.20 \\ -4.58 &3.07 &-2.74 &3292.41 &2.43 \\ -13.64& 9.21 &-8.20 &2.43 &3292.67 \end{pmatrix} $}  &2764.52 & (\textbf{0.99963}, -0.00061, 0.00041, 0.00856, 0.02579)\\
     & $(0,0)2S(\frac{3}{2}^{+})_{1}$ & &3196.72 &(-0.00331, -\textbf{0.99503}, 0.00602, 0.02953, 0.09481)\\
      & $(0,0)3S(\frac{3}{2}^{+})_{1}$ & &3290.09 &(-0.01170, 0.04507, 0.07525, -\textbf{0.71717}, 0.69126)\\
       & $(0,2)1D(\frac{3}{2}^{+})_{1}$ & &3294.72 &(-0.02387, 0.08785, 0.17033, 0.69314, \textbf{0.69445})\\
    & $(0,2)1D(\frac{3}{2}^{+})_{2}$ & &3340.93 &(-0.00464, 0.01259, -\textbf{0.98249}, 0.06542, 0.17393)\\\hline
    \multirow{2}{*}{$\Omega_{c}(\frac{5}{2}^{+})$} &$(0,2)1D(\frac{5}{2}^{+})_{2}$ & \multirow{2}{*}{$ \begin{pmatrix} 3311.34 & -2.27 \\ -2.27 & 3309.79 \end{pmatrix} $}  &3308.17 & (-0.581765, -\textbf{0.813357})\\
     & $(0,2)1D(\frac{5}{2}^{+})_{3}$ & &3312.96 &(-\textbf{0.813357}, 0.581765)\\\hline
    $\Omega_{c}(\frac{7}{2}^{+})$ & $(0,2)1D(\frac{7}{2}^{+})_{3}$ & (3332.23) & 3332.23 & (1) \\
\end{tabular}
\end{ruledtabular}
\end{table*}

\begin{table*}[htbp]
\begin{ruledtabular}\caption{Energy level shifts caused by the mixing effect and the contribution of each spin-dependent term to the mixed mass $M_{mixed}$ for the positive-parity charm baryons. $M_{mixed}$ is the mass (eigen-energy) of the baryon state as shown in Tables~\ref{tb5} and~\ref{tb6}. $M_{unmixed}$ is the energy level of the special orbitally excited state defined in Table~\ref{tb5}, the main contributor to the baryon state. $\Delta M$ = $M_{mixed}$ - $M_{unmixed}$ is the mass shift of the special orbitally excited state in the mixing effect. $M_{mixed}$ is composed of the diagonal matrix $C^{\dag}H^{(0)}C$ and six spin-dependent terms being considered in the mixing effect $C^{\dag}H^{(i)}C$ ($i\neq0$), i.e., $M_{mixed}$ = $C^{\dag}H^{(0)}C$ + $\sum_{i=1}^{6}C^{\dag}H^{(i)}C$. Here, $C$ is the eigenvector of each eigen-energy $M_{mixed}$ (see Tables~\ref{tb5} and~\ref{tb6}). $H^{(i)}$ denotes the matrix of each spin-dependent term (the only source of the configuration mixing) and can be obtained from Tables~\ref{tb3} and~\ref{tb4}, and its diagonal elements are all zero. $H^{(0)}$ is a diagonal matrix keeping only the diagonal elements of $H$ in Tables~\ref{tb5} and~\ref{tb6}.}
\begin{tabular}{c c c c c c c c c c c c c}
\label{tb7}
\multirow{2}{*}{$J^{P}$} & \multirow{2}{*}{Orbital State} & \multirow{2}{*}{$M_{unmixed}$} & \multirow{2}{*}{$M_{mixed}$} & \multirow{2}{*}{$\Delta M$} & \multicolumn{7}{c}{$C^{\dag}H^{(i)}C$} & \multirow{2}{*}{$C^{\dag}H^{(0)}C$}\\ \cline{6-12}
 & & & & & $\hat{H}^{tens}$ & $\hat{H}^{cont}$ & $\hat{H}^{so(v)-2B}$  & $\hat{H}^{so(s)-2B}$ & $\hat{H}^{so(s)-3B}$  & $\hat{H}^{so(v)-3B}$ & Sum \\ \hline
\multirow{4}{*}{$\Sigma_{c}(\frac{1}{2}^{+})$} &$(0,0)1S(\frac{1}{2}^{+})_{1}$ & 2456.24 & 2456.10 &-0.14 &-0.278&0&0&0&0&0&-0.278&2456.38 \\
&  $(0,0)2S(\frac{1}{2}^{+})_{1}$ & 2913.44 &2913.05 &-0.39&-0.783&0&0&0&0&0&-0.783&2913.83\\
& $(0,0)3S(\frac{1}{2}^{+})_{1}$ & 3109.47 &3109.75 &0.28&0.569&0&0&0&0&0&0.569&3109.18\\
& $(0,2)1D(\frac{1}{2}^{+})_{1}$ & 3048.14 & 3048.39& 0.25& 0.492&0&0&0&0&0& 0.492&3047.90\\ \hline
\multirow{5}{*}{$\Sigma_{c}(\frac{3}{2}^{+})$} &$(0,0)1S(\frac{3}{2}^{+})_{1}$ & 2533.92 & 2533.51&-0.41 & -0.816&-0.002&-0.005&0.002&-0.000&0.001&-0.820&2534.33\\
 & $(0,0)2S(\frac{3}{2}^{+})_{1}$ & 2967.28 & 2966.33 &-0.95&-1.825&-0.026&-0.063&0.025&-0.002&0.018&-1.873&2968.20\\
  & $(0,0)3S(\frac{3}{2}^{+})_{1}$ & 3126.65 & 3127.48 &0.83&1.730&-0.037&-0.089&0.036&-0.003&0.025&1.662&3125.81\\
   & $(0,2)1D(\frac{3}{2}^{+})_{1}$ & 3062.98 & 3063.69& 0.71& 8.997&-4.126&-9.942&3.966&-0.302&2.792&1.393&3062.29\\
   & $(0,2)1D(\frac{3}{2}^{+})_{2}$ &3061.57 & 3061.40 &-0.17&-8.086&4.192&10.100&-4.029&0.307&-2.837&-0.362&3061.76\\ \hline
   \multirow{2}{*}{$\Sigma_{c}(\frac{5}{2}^{+})$} & $(0,2)1D(\frac{5}{2}^{+})_{2}$ & 3082.51 & 3082.76 & 0.25&2.818&-1.260&-3.265&1.394&-0.102&0.892& 0.486&3082.28\\
   & $(0,2)1D(\frac{5}{2}^{+})_{3}$ &3076.68 & 3076.43 & -0.25&-2.818&1.260&3.265&-1.394&0.102&-0.892&-0.485 &3076.91\\ \hline \hline
 \multirow{4}{*}{$\Xi'_{c}(\frac{1}{2}^{+})$} &$(0,0)1S(\frac{1}{2}^{+})_{1}$ & 2589.18 &2589.06 &-0.12&-0.242&0&0&0&0&0&-0.242&2589.30 \\
 & $(0,0)2S(\frac{1}{2}^{+})_{1}$ & 3046.19 &3045.73&-0.46&-0.911&0&0&0&0&0&-0.911&3046.64\\
 & $(0,0)3S(\frac{1}{2}^{+})_{1}$ & 3219.50 & 3219.70&0.20&0.395&0&0&0&0&0&0.395&3219.30\\
 & $(0,2)1D(\frac{1}{2}^{+})_{1}$ & 3176.65 & 3177.03&0.38&0.758&0&0&0&0&0&0.758&3176.27\\ \hline
\multirow{5}{*}{$\Xi'_{c}(\frac{3}{2}^{+})$} &$(0,0)1S(\frac{3}{2}^{+})_{1}$ & 2660.13 & 2659.77 &-0.36&-0.723&-0.002&-0.004&0.002&-0.000&0.001&-0.726&2660.49\\
& $(0,0)2S(\frac{3}{2}^{+})_{1}$ &3095.71 & 3094.56 &-1.15&-2.191&-0.028&-0.069&0.030&-0.002&0.015&-2.245&3096.81\\
 & $(0,0)3S(\frac{3}{2}^{+})_{1}$ &3237.15 & 3237.79&0.64&1.378&-0.037&-0.091&0.040&-0.003&0.021&1.307&3236.48\\
  & $(0,2)1D(\frac{3}{2}^{+})_{1}$ &3189.00 & 3187.54&-1.46&-9.089&3.868&9.402&-4.116&0.313&-2.118&-1.741&3189.28\\
  & $(0,2)1D(\frac{3}{2}^{+})_{2}$ &3190.16 & 3192.49& 2.33&10.625&-3.800&-9.238&4.045&-0.308&2.081&3.405&3189.08\\ \hline
  \multirow{2}{*}{$\Xi'_{c}(\frac{5}{2}^{+})$} &$(0,2)1D(\frac{5}{2}^{+})_{2}$ &3208.31 & 3210.05&1.74&7.643&-2.832&-7.283&3.400&-0.237&1.582&2.292&3207.75\\
   & $(0,2)1D(\frac{5}{2}^{+})_{3}$ &3206.67 & 3204.93& -1.74&-7.643&2.832&7.283&-3.400&0.237&-1.582&-2.292&3207.23\\ \hline \hline
 \multirow{4}{*}{$\Omega_{c}(\frac{1}{2}^{+})$} &$(0,0)1S(\frac{1}{2}^{+})_{1}$ & 2696.35 &2696.21&-0.14&-0.272&0&0&0&0&0&-0.272&2696.49\\
 & $(0,0)2S(\frac{1}{2}^{+})_{1}$ & 3150.05 & 3149.61&-0.44&-0.873&0&0&0&0&0&-0.873&3150.48\\
  & $(0,0)3S(\frac{1}{2}^{+})_{1}$ & 3324.71 & 3325.28& 0.57&1.146&0&0&0&0&0&1.146&3324.14\\
  & $(0,2)1D(\frac{1}{2}^{+})_{1}$ & 3278.33 & 3278.33&0.00&-0.001&0&0&0&0&0&-0.001&3278.33\\ \hline
  \multirow{5}{*}{$\Omega_{c}(\frac{3}{2}^{+})$} &$(0,0)1S(\frac{3}{2}^{+})_{1}$ &2764.91 &2764.52&-0.39&-0.778&-0.002&-0.005&0.000&-0.000&0.001&-0.781&2765.30\\
   & $(0,0)2S(\frac{3}{2}^{+})_{1}$ & 3197.69 &3196.72&-0.87&-1.866&-0.020&-0.060&0.004&-0.002&0.017&-1.905&3198.63\\
   & $(0,0)3S(\frac{3}{2}^{+})_{1}$ &3339.30 &3340.93&1.63&3.439&-0.081&-0.244&0.015&-0.008&0.068&3.280&3337.65\\
    & $(0,2)1D(\frac{3}{2}^{+})_{1}$ &3292.41 &3290.09 &-2.32&-9.358&3.530&10.609&-0.635&0.337&-2.974&-2.447&3292.54\\
   & $(0,2)1D(\frac{3}{2}^{+})_{2}$ &3292.67 &3294.72&2.05 &8.564&-3.427&-10.301&0.616&-0.327&2.888&1.854&3292.86\\ \hline
   \multirow{2}{*}{$\Omega_{c}(\frac{5}{2}^{+})$} &$(0,2)1D(\frac{5}{2}^{+})_{2}$ & 3311.34 &3312.96&1.62&6.776&-2.384&-7.590&0.473&-0.246&2.063&2.148&3310.82\\
    & $(0,2)1D(\frac{5}{2}^{+})_{3}$ & 3309.79 &3308.17&-1.62&-6.776&2.385&7.590&-0.473&0.246&-2.063&-2.148&3310.31\\
\end{tabular}
\end{ruledtabular}
\end{table*}

\begin{table*}[htbp]
\begin{ruledtabular}\caption{Same as Table~\ref{tb7}, but for the positive-parity bottom baryons.}
\begin{tabular}{c c c c c c c c c c c c c}
\label{tb8}
\multirow{2}{*}{$J^{P}$} & \multirow{2}{*}{Orbital State} & \multirow{2}{*}{$M_{unmixed}$} & \multirow{2}{*}{$M_{mixed}$} & \multirow{2}{*}{$\Delta M$} & \multicolumn{7}{c}{$C^{\dag}H^{(i)}C$} & \multirow{2}{*}{$C^{\dag}H^{(0)}C$}\\ \cline{6-12}
 & & & & & $\hat{H}^{tens}$ & $\hat{H}^{cont}$ & $\hat{H}^{so(v)-2B}$  & $\hat{H}^{so(s)-2B}$ & $\hat{H}^{so(s)-3B}$  & $\hat{H}^{so(v)-3B}$ & Sum \\ \hline
\multirow{4}{*}{$\Sigma_{b}(\frac{1}{2}^{+})$} &$(0,0)1S(\frac{1}{2}^{+})_{1}$ & 5820.75 & 5820.73 &-0.02 &-0.048&0&0&0&0&0&-0.048&5820.77  \\
&  $(0,0)2S(\frac{1}{2}^{+})_{1}$ & 6226.18 &6226.12 &-0.06&-0.116&0&0&0&0&0&-0.116&6226.24 \\
& $(0,0)3S(\frac{1}{2}^{+})_{1}$ & 6438.63 &6438.65 &0.02&0.048&0&0&0&0&0&0.048&6438.61\\
& $(0,2)1D(\frac{1}{2}^{+})_{1}$ & 6329.76 & 6329.82& 0.06& 0.116&0&0&0&0&0& 0.116&6329.70 \\ \hline
\multirow{5}{*}{$\Sigma_{b}(\frac{3}{2}^{+})$} &$(0,0)1S(\frac{3}{2}^{+})_{1}$ & 5849.15 & 5849.09&-0.06 & -0.125&-0.000&-0.000&0.000&-0.000&0.000&-0.125&5849.21 \\
 & $(0,0)2S(\frac{3}{2}^{+})_{1}$ & 6245.90 & 6245.76 &-0.14&-0.282&-0.002&-0.004&0.001&-0.000&0.001&-0.286&6246.04 \\
  & $(0,0)3S(\frac{3}{2}^{+})_{1}$ & 6446.14 & 6446.20 &0.06&0.123&-0.001&-0.001&0.000&-0.000&0.000&0.121&6446.08\\
   & $(0,2)1D(\frac{3}{2}^{+})_{1}$ & 6337.32 & 6337.40& 0.08& -1.087&0.599&1.107&-0.164&0.011&-0.328&0.144&6337.25 \\
   & $(0,2)1D(\frac{3}{2}^{+})_{2}$ &6334.23 & 6334.30 &0.07&1.371&-0.596&-1.101&0.163&-0.011&0.326&0.146&6334.15 \\ \hline
   \multirow{2}{*}{$\Sigma_{b}(\frac{5}{2}^{+})$} & $(0,2)1D(\frac{5}{2}^{+})_{2}$ & 6344.72 & 6344.72 & 0.00&-0.121&0.060&0.118&-0.019&0.002&-0.034& 0.006&6344.72\\
   & $(0,2)1D(\frac{5}{2}^{+})_{3}$ &6337.61 & 6337.61 & 0.00&0.121&-0.060&-0.118&0.019&-0.002&0.034&-0.006 &6337.61 \\ \hline \hline
 \multirow{4}{*}{$\Xi'_{b}(\frac{1}{2}^{+})$} &$(0,0)1S(\frac{1}{2}^{+})_{1}$ & 5944.49 &5944.47 &-0.02&-0.040&0&0&0&0&0&-0.040&5944.51  \\
 & $(0,0)2S(\frac{1}{2}^{+})_{1}$ & 6351.02 &6350.95&-0.07&-0.146&0&0&0&0&0&-0.146&6351.09 \\
 & $(0,0)3S(\frac{1}{2}^{+})_{1}$ & 6542.97 & 6542.99&0.02&0.034&0&0&0&0&0&0.034&6542.95\\
 & $(0,2)1D(\frac{1}{2}^{+})_{1}$ & 6451.54 & 6451.62&0.08&0.152&0&0&0&0&0&0.152&6451.46 \\ \hline
\multirow{5}{*}{$\Xi'_{b}(\frac{3}{2}^{+})$} &$(0,0)1S(\frac{3}{2}^{+})_{1}$ & 5970.89 & 5970.84 &-0.05&-0.107&-0.000&-0.000&0.000&-0.000&0.000&-0.107&5970.94 \\
& $(0,0)2S(\frac{3}{2}^{+})_{1}$ &6369.31 & 6369.13 &-0.18&-0.363&-0.002&-0.004&0.001&-0.000&0.001&-0.368&6369.49 \\
 & $(0,0)3S(\frac{3}{2}^{+})_{1}$ &6550.94 & 6550.98&0.04&0.090&-0.001&-0.001&0.000&-0.000&0.000&0.089&6550.90\\
  & $(0,2)1D(\frac{3}{2}^{+})_{1}$ &6458.27 & 6458.29&0.02&0.067&-0.013&-0.025&0.004&-0.000&0.006&0.038&6458.25 \\
  & $(0,2)1D(\frac{3}{2}^{+})_{2}$ &6456.23 & 6456.40& 0.17&0.313&0.016&0.031&-0.005&0.000&-0.007&0.348&6456.06 \\ \hline
  \multirow{2}{*}{$\Xi'_{b}(\frac{5}{2}^{+})$} &$(0,2)1D(\frac{5}{2}^{+})_{2}$ &6465.73 & 6465.74&0.01&0.374&-0.156&-0.304&0.054&-0.004&0.067&0.030&6465.72\\
   & $(0,2)1D(\frac{5}{2}^{+})_{3}$ &6460.84 & 6460.83& -0.01&-0.374&0.156&0.304&-0.054&0.004&-0.067&-0.030&6460.85\\ \hline \hline
 \multirow{4}{*}{$\Omega_{b}(\frac{1}{2}^{+})$} &$(0,0)1S(\frac{1}{2}^{+})_{1}$ & 6043.10 &6043.07&-0.03&-0.052&0&0&0&0&0&-0.052&6043.13 \\
 & $(0,0)2S(\frac{1}{2}^{+})_{1}$ & 6447.79 & 6447.72&-0.07&-0.150&0&0&0&0&0&-0.150&6447.86 \\
  & $(0,0)3S(\frac{1}{2}^{+})_{1}$ & 6640.56 & 6640.61& 0.05&0.093&0&0&0&0&0&0.093&6640.51\\
  & $(0,2)1D(\frac{1}{2}^{+})_{1}$ & 6544.34 & 6544.39&0.05&0.108&0&0&0&0&0&0.108&6544.29 \\ \hline
  \multirow{5}{*}{$\Omega_{b}(\frac{3}{2}^{+})$} &$(0,0)1S(\frac{3}{2}^{+})_{1}$ &6069.03 &6068.96&-0.07&-0.136&-0.000&-0.000&0.000&-0.000&0.000&-0.136&6069.10 \\
   & $(0,0)2S(\frac{3}{2}^{+})_{1}$ & 6465.47 &6465.30&-0.17&-0.339&-0.002&-0.005&0.001&-0.000&0.001&-0.343&6465.64 \\
   & $(0,0)3S(\frac{3}{2}^{+})_{1}$ &6647.24 &6647.36&0.12&0.239&-0.001&-0.003&0.000&-0.000&0.001&0.236&6647.12\\
    & $(0,2)1D(\frac{3}{2}^{+})_{1}$ &6551.96 &6552.08 &0.12&2.187&-0.862&-1.934&0.321&-0.030&0.581&0.253&6551.83 \\
   & $(0,2)1D(\frac{3}{2}^{+})_{2}$ &6550.50 &6550.50&0.00 &-1.951&0.865&1.942&-0.322&0.030&-0.584&-0.009&6550.51 \\ \hline
   \multirow{2}{*}{$\Omega_{b}(\frac{5}{2}^{+})$} &$(0,2)1D(\frac{5}{2}^{+})_{2}$ & 6560.51 &6560.58&0.07&0.934&-0.354&-0.818&0.138&-0.011&0.243&0.135&6560.44\\
    & $(0,2)1D(\frac{5}{2}^{+})_{3}$ & 6557.10 &6557.03&-0.07&-0.934&0.354&0.818&-0.138&0.011&-0.243&-0.135&6557.17\\
\end{tabular}
\end{ruledtabular}
\end{table*}

\begin{table*}[htbp]
\begin{ruledtabular}\caption{Same as Table~\ref{tb7}, but for the negative-parity singly heavy baryons. The original data used are from Ref.~\cite{Li26}.}
\begin{tabular}{c c c c c c c c c c c c c}
\label{tb9}
\multirow{2}{*}{$J^{P}$} & \multirow{2}{*}{Orbital State} & \multirow{2}{*}{$M_{unmixed}$} & \multirow{2}{*}{$M_{mixed}$} & \multirow{2}{*}{$\Delta M$} & \multicolumn{7}{c}{$C^{\dag}H^{(i)}C$} & \multirow{2}{*}{$C^{\dag}H^{(0)}C$}\\ \cline{6-12}
 & & & & & $\hat{H}^{tens}$ & $\hat{H}^{cont}$ & $\hat{H}^{so(v)-2B}$  & $\hat{H}^{so(s)-2B}$ & $\hat{H}^{so(v)-3B}$  & $\hat{H}^{so(s)-3B}$ & Sum \\ \hline
\multirow{3}{*}{$\Sigma_{c}(\frac{1}{2}^{-})$} &$(0,1)1P(\frac{1}{2}^{-})_{0}$ & 2773.06 & 2783.20 &10.14 &-9.263&12.834&16.607&-3.931&1.611&-12.818&5.026&2778.17   \\
&  $(0,1)1P(\frac{1}{2}^{-})_{1}$ & 2778.01 &2759.41 &-18.60&4.892&-5.790&-26.284&6.422&18.079&-17.255&-19.925&2779.33  \\
& $(1,0)1P(\frac{1}{2}^{-})_{1}$ & 2828.13 &2836.60 &8.47&4.372&-7.043&9.678&-2.491&-19.689&30.072&14.899&2821.71  \\ \hline
\multirow{3}{*}{$\Sigma_{c}(\frac{3}{2}^{-})$} &$(0,1)1P(\frac{3}{2}^{-})_{1}$ & 2810.40& 2798.58&-11.82 &11.135&-10.824&-17.390&4.530&18.390&-24.358&-18.525&2817.10  \\
 & $(0,1)1P(\frac{3}{2}^{-})_{2}$ & 2816.13& 2818.41 &2.28&-8.341&8.217&8.291&-2.099&-0.536&-4.145&1.406&2817.00  \\
  & $(1,0)1P(\frac{3}{2}^{-})_{1}$ & 2877.37 & 2886.91 &9.54&-2.794&2.607&9.099&-2.431&-17.854&28.504&17.118&2869.79 \\ \hline \hline
\multirow{3}{*}{$\Xi'_{c}(\frac{1}{2}^{-})$} &$(0,1)1P(\frac{1}{2}^{-})_{0}$ & 2906.46 & 2895.81 &-10.65 &7.130&-7.147&-23.581&6.132&13.234&-10.466&-14.714&2910.52   \\
&  $(0,1)1P(\frac{1}{2}^{-})_{1}$ & 2911.67 &2918.64 &6.97&-11.021&12.119&17.470&-4.485&3.463&-9.637&7.931&2910.71  \\
& $(1,0)1P(\frac{1}{2}^{-})_{1}$ & 2958.12 &2961.80 &3.68&3.891&-4.972&6.111&-1.647&-16.697&20.103&6.784&2955.02  \\ \hline
\multirow{3}{*}{$\Xi'_{c}(\frac{3}{2}^{-})$} &$(0,1)1P(\frac{3}{2}^{-})_{1}$ & 2940.83 & 2934.03&-6.8 &10.538&-9.106&-15.691&4.324&13.814&-14.268&-10.391&2944.42  \\
 & $(0,1)1P(\frac{3}{2}^{-})_{2}$ & 2947.58& 2950.25 &2.67&-9.079&7.915&9.569&-2.607&1.185&-4.536&2.444&2947.81  \\
  & $(1,0)1P(\frac{3}{2}^{-})_{1}$ & 3004.10 & 3008.23 &4.13&-1.459&1.190&6.122&-1.717&-14.999&18.804&7.946&3000.28 \\ \hline \hline
\multirow{3}{*}{$\Omega_{c}(\frac{1}{2}^{-})$} &$(0,1)1P(\frac{1}{2}^{-})_{0}$ & 3009.38& 3001.40 &-7.98&6.564&-6.046&-24.283&6.079&13.487&-7.850&-12.018&3013.42   \\
&  $(0,1)1P(\frac{1}{2}^{-})_{1}$ & 3014.79 &3019.70 &4.91&-11.914&11.464&20.278&-5.131&-3.131&-5.103&6.466&3013.23  \\
& $(1,0)1P(\frac{1}{2}^{-})_{1}$ & 3058.84 &3061.91 &3.07&5.350&-5.418&4.005&-0.947&-10.355&12.954&5.552&3056.36  \\ \hline
\multirow{3}{*}{$\Omega_{c}(\frac{3}{2}^{-})$} &$(0,1)1P(\frac{3}{2}^{-})_{1}$ & 3044.57 & 3040.85&-3.72 &7.349&-6.775&-12.908&3.404&10.116&-8.958&-6.052&3046.90  \\
 & $(0,1)1P(\frac{3}{2}^{-})_{2}$ & 3051.82& 3052.59 &0.77&-9.349&6.480&8.447&-2.279&-1.259&-2.278&0.263&3052.33  \\
  & $(1,0)1P(\frac{3}{2}^{-})_{1}$ & 3108.83 & 3111.78 &2.95&2.000&0.295&4.461&-1.125&-8.856&11.236&5.790&3105.99 \\ \hline \hline
\multirow{2}{*}{$\Sigma_{b}(\frac{1}{2}^{-})$} &$(0,1)1P(\frac{1}{2}^{-})_{0}$ & 6087.11 & 6084.04 &-3.07 &4.017&-4.831&-6.530&0.619&2.301&-0.071&-4.495&6088.54   \\
&  $(0,1)1P(\frac{1}{2}^{-})_{1}$ & 6092.46 &6095.53 &3.07&-4.017&4.831&6.530&-0.619&-2.301&0.071&4.495&6091.03  \\ \hline
\multirow{2}{*}{$\Sigma_{b}(\frac{3}{2}^{-})$} &$(0,1)1P(\frac{3}{2}^{-})_{1}$ & 6105.46 & 6104.22&-1.24 &2.352&-2.508&-3.553&0.353&1.254&-0.039&-2.175&6106.40  \\
 & $(0,1)1P(\frac{3}{2}^{-})_{2}$ & 6113.18& 6114.42 &1.24&-2.352&2.508&3.553&-0.353&-1.254&0.039&2.175&6112.24  \\ \hline \hline

\multirow{2}{*}{$\Xi'_{b}(\frac{1}{2}^{-})$} &$(0,1)1P(\frac{1}{2}^{-})_{0}$ & 6213.75 & 6210.71 &-3.04 &4.138&-4.300&-6.297&0.648&1.502&-0.063&-4.372&6215.08   \\
&  $(0,1)1P(\frac{1}{2}^{-})_{1}$ & 6218.47 &6221.51 &3.04&-4.138&4.300&6.297&-0.648&-1.502&0.063&4.372&6217.14  \\ \hline
\multirow{2}{*}{$\Xi'_{b}(\frac{3}{2}^{-})$} &$(0,1)1P(\frac{3}{2}^{-})_{1}$ & 6230.38  & 6229.07&-1.31 &2.447&-2.351&-3.551&0.377&0.843&-0.041&-2.269&6231.34  \\
 & $(0,1)1P(\frac{3}{2}^{-})_{2}$ & 6237.41 &6238.72 &1.31&-2.447&2.351&3.551&-0.377&-0.843&0.041&2.269&6236.45  \\ \hline \hline
\multirow{2}{*}{$\Omega_{b}(\frac{1}{2}^{-})$} &$(0,1)1P(\frac{1}{2}^{-})_{0}$ & 6307.76 &6305.88 &-1.88 &4.041&-3.720&-6.078&0.609&2.197&-0.064&-3.007&6308.88   \\
&  $(0,1)1P(\frac{1}{2}^{-})_{1}$ & 6313.35 &6315.23 &1.88&-4.041&3.720&6.078&-0.609&-2.197&0.064&3.007&6312.23  \\ \hline
\multirow{2}{*}{$\Omega_{b}(\frac{3}{2}^{-})$} &$(0,1)1P(\frac{3}{2}^{-})_{1}$ & 6325.67 & 6325.16&-0.51 &1.857&-1.475&-2.488&0.258&0.898&-0.027&-0.969&6326.13  \\
 & $(0,1)1P(\frac{3}{2}^{-})_{2}$ & 6334.46& 6334.97 &0.51&-1.857&1.475&2.488&-0.258&-0.898&0.027&0.969&6334.00  \\
\end{tabular}
\end{ruledtabular}
\end{table*}

\begin{table*}[htbp]
\begin{ruledtabular}\caption{The assignment of the 74 observed baryons in this work. The first two columns are from experimental data~\cite{F201,F209,LHCb25,LHCb26}. The third and fourth columns are the calculation data and assignments in this work. The deviations of the calculated masses from the measured ones are listed in the fifth column. Most of the absolute deviations are less than 16 MeV.  The maximum absolute deviation is 18.70 MeV. The numbers of states lying within the absolute deviation  intervals [$0\sim5$), [$5\sim10$), [$10\sim15$) and [$15\sim20$) MeV are 39, 13, 13 and 9, respectively. In the 74 samples of $\{(M_{cal.}- M_{exp.})_{i}\}$, the sample median and the median absolute deviation are 0.75 and 5.13 MeV, respectively. The mean absolute deviation (MAD) $(\sum_{i=1}^{n}|M_{cal.}-M_{exp.}|_{i})/n$ is 6.88 MeV. And the root mean square deviation (RMSD) $\sqrt{\frac{1}{n}\sum_{i=1}^{n}(M_{cal.}-M_{exp.})_{i}^{2}}$ is 8.71 MeV. $M_{exp.}$ denotes the central value of the measured mass. `$\uparrow$' means the same as above. $\Lambda_{c}(2940)^{+}$, $\Lambda_{b}(6070)^{0}$ and $\Xi_{c}(3123)^{+}$ are not included in the list. }
\begin{tabular}{c c c c c | c c c c c }
\label{tb10}
Baryon ($J^{P}$) & $M_{exp.}$  & Baryon $(mass)J^{P}$ & $M_{cal.}$ & $M_{cal.}$-$M_{exp.}$  & Baryon ($J^{P}$) & $M_{exp.}$  & Baryon $(mass)J^{P}$ & $M_{cal.}$ & $M_{cal.}$-$M_{exp.}$ \\ \hline
$\Lambda_{c}^{+}(\frac{1}{2}^{+})$ & 2286.46  & $\Lambda_{c}(2288)\frac{1}{2}^{+}$ & 2288 & 1.54 & $\Omega_{c}(2770)^{0}(\frac{3}{2}^{+})$ & 2766  & $\Omega_{c}(2765)\frac{3}{2}^{+}$ & 2765 & -1.00 \\
$\Lambda_{c}(2595)^{+}(\frac{1}{2}^{-})$ & 2592.25 & $\Lambda_{c}(2597)\frac{1}{2}^{-}$ & 2597 & 4.75& $\Omega_{c}(3000)^{0}(?^{?})$ & 3000.46  & $\Omega_{c}(3001)\frac{1}{2}^{-}$ & 3001 & 0.54  \\
$\Lambda_{c}(2625)^{+}(\frac{3}{2}^{-})$ & 2628  & $\Lambda_{c}(2631)\frac{3}{2}^{-}$ & 2631 & 3.00 & $\Omega_{c}(3050)^{0}(?^{?})$ & 3050.17  & $\Omega_{c}(3053)\frac{3}{2}^{-}$ & 3053 & 2.83  \\
$\Lambda_{c}(2765)^{+}(?^{?})$ & 2766.6  & $\Lambda_{c}(2764)\frac{1}{2}^{+}$ & 2764 & -2.60 & $\Omega_{c}(3065)^{0}(?^{?})$ & 3065.58  & $\Omega_{c}(3062)\frac{1}{2}^{-}$ & 3062 & -3.58  \\
$\Lambda_{c}(2860)^{+}(\frac{3}{2}^{+})$ & 2856.1  & $\Lambda_{c}(2873)\frac{3}{2}^{+}$ & 2873 & 16.90 & $\Omega_{c}(3090)^{0}(?^{?})$ & 3090.15  & $\Omega_{c}(3095)\frac{5}{2}^{-}$ & 3095 & 4.85  \\
$\Lambda_{c}(2880)^{+}(\frac{5}{2}^{+})$ & 2881.62  & $\Lambda_{c}(2892)\frac{5}{2}^{+}$ & 2892 & 10.38 & $\Omega_{c}(3120)^{0}(?^{?})$ & 3118.98  & $\Omega_{c}(3112)\frac{3}{2}^{-}$ & 3112 & -6.98  \\
$\Sigma_{c}(2455)^{++}(\frac{1}{2}^{+})$ & 2453.97  & $\Sigma_{c}(2456)\frac{1}{2}^{+}$ & 2456 & 2.03 & $\Omega_{c}(3185)^{0}(?^{?})$ & 3185  & $\Omega_{c}(3197)\frac{3}{2}^{+}$ & 3197 & 12.00  \\
$\Sigma_{c}(2455)^{+}(\frac{1}{2}^{+})$ & 2452.65  & $\uparrow$ & $\uparrow$ & 3.35 & $\Omega_{c}(3327)^{0}(?^{?})$ & 3327.1  & $\Omega_{c}(3325)\frac{1}{2}^{+}$ & 3325 & -2.10  \\
$\Sigma_{c}(2455)^{0}(\frac{1}{2}^{+})$ & 2453.75  & $\uparrow$ & $\uparrow$ & 2.25  & $\Lambda_{c}(2910)^{0}(?^{?})$ & 2914  & $\Sigma_{c}(2913)\frac{1}{2}^{+}$ & 2913 & -1.00 \\
$\Sigma_{c}(2520)^{++}(\frac{3}{2}^{+})$ & 2518.42  & $\Sigma_{c}(2534)\frac{3}{2}^{+}$ & 2534 & 15.58 & $\Lambda_{b}^{0}(\frac{1}{2}^{+})$ & 5619.5  & $\Lambda_{b}(5622)\frac{1}{2}^{+}$ & 5622 & 2.43 \\
$\Sigma_{c}(2520)^{+}(\frac{3}{2}^{+})$ & 2517.4  & $\uparrow$ & $\uparrow$ & 16.60 & $\Lambda_{b}(5912)^{0}(\frac{1}{2}^{-})$ & 5912.16  & $\Lambda_{b}(5899)\frac{1}{2}^{-}$ & 5899 & -13.20 \\
$\Sigma_{c}(2520)^{0}(\frac{3}{2}^{+})$ & 2518.48  & $\uparrow$ & $\uparrow$ & 15.52 & $\Lambda_{b}(5920)^{0}(\frac{3}{2}^{-})$ & 5920.07  & $\Lambda_{b}(5913)\frac{3}{2}^{-}$ & 5913 & -7.07 \\
$\Sigma_{c}(2800)^{++}(?^{?})$ & 2801  & $\Sigma_{c}(2799)\frac{3}{2}^{-}$ & 2799 & -2.00& $\Lambda_{b}(6146)^{0}(\frac{3}{2}^{+})$ & 6146.2  & $\Lambda_{b}(6135)\frac{3}{2}^{+}$ & 6135 & -11.20 \\
$\Sigma_{c}(2800)^{+}(?^{?})$ & 2792  & $\uparrow$ & $\uparrow$ & 7.00 & $\Lambda_{b}(6152)^{0}(\frac{5}{2}^{+})$ & 6152.5  & $\Lambda_{b}(6146)\frac{5}{2}^{+}$ & 6146 & -6.50 \\
$\Sigma_{c}(2800)^{0}(?^{?})$ & 2806  & $\uparrow$ & $\uparrow$ & -7.00 &$\Sigma_{b}^{+}(\frac{1}{2}^{+})$   & 5810.56  & $\Sigma_{b}(5821)\frac{1}{2}^{+}$ & 5821 & 10.44 \\
$\Sigma_{c}(2846)^{0}(?^{?})$ & 2846  & $\Sigma_{c}(2837)\frac{1}{2}^{-}$ & 2837 & -9.00 & $\Sigma_{b}^{-}(\frac{1}{2}^{+})$ & 5815.64 & $\uparrow$ & $\uparrow$ & 5.36 \\
$\Xi_{c}^{+}(\frac{1}{2}^{+})$ & 2467.95  & $\Xi_{c}(2479)\frac{1}{2}^{+}$ & 2479 & 11.05 & $\Sigma_{b}^{*+}(\frac{3}{2}^{+})$ & 5830.32 & $\Sigma_{b}(5849)\frac{3}{2}^{+}$ & 5849 & 18.68 \\
$\Xi_{c}^{0}(\frac{1}{2}^{+})$ & 2470.44  & $\uparrow$ & $\uparrow$ & 8.56 & $\Sigma_{b}^{*-}(\frac{3}{2}^{+})$ & 5834.74 & $\uparrow$ & $\uparrow$ & 14.26 \\
$\Xi_{c}^{'+}(\frac{1}{2}^{+})$ & 2578.2  & $\Xi'_{c}(2589)\frac{1}{2}^{+}$ & 2589 & 10.80 & $\Sigma_{b}(6097)^{+}(?^{?})$ & 6095.8 & $\Sigma_{b}(6096)\frac{1}{2}^{-}$ & 6096 & 0.20 \\
$\Xi_{c}^{'0}(\frac{1}{2}^{+})$ & 2578.7  & $\uparrow$ & $\uparrow$ & 10.30 & $\Sigma_{b}(6097)^{-}(?^{?})$ & 6098.0 & $\uparrow$ & $\uparrow$ & -2.00 \\
$\Xi_{c}(2645)^{+}(\frac{3}{2}^{+})$ & 2645.1  & $\Xi'_{c}(2660)\frac{3}{2}^{+}$ & 2660 & 14.90 & $\Xi_{b}^{-}(\frac{1}{2}^{+})$  & 5797 & $\Xi_{b}(5806)\frac{1}{2}^{+}$ & 5806 & 9.00 \\
$\Xi_{c}(2645)^{0}(\frac{3}{2}^{+})$ & 2645.7  & $\uparrow$ & $\uparrow$ & 14.30 & $\Xi_{b}^{0}(\frac{1}{2}^{+})$  & 5791.7 & $\uparrow$ & $\uparrow$ & 14.30 \\
$\Xi_{c}(2790)^{+}(\frac{1}{2}^{-})$ & 2791.9  & $\Xi_{c}(2789)\frac{1}{2}^{-}$ & 2789 & -2.90 & $\Xi_{b}(5935)^{-}(\frac{1}{2}^{+})$  & 5934.9 & $\Xi'_{b}(5944)\frac{1}{2}^{+}$ & 5944 & 9.10 \\
$\Xi_{c}(2790)^{0}(\frac{1}{2}^{-})$ & 2793.9  & $\uparrow$ & $\uparrow$ & -4.90 & $\Xi_{b}(5945)^{0}(\frac{3}{2}^{+})$  & 5952.3 & $\Xi'_{b}(5971)\frac{3}{2}^{+}$ & 5971 & 18.70 \\
$\Xi_{c}(2815)^{+}(\frac{3}{2}^{-})$ & 2816.51  & $\Xi_{c}(2820)\frac{3}{2}^{-}$ & 2820 & 3.49 & $\Xi_{b}(5955)^{-}(\frac{3}{2}^{+})$  & 5955.5 & $\uparrow$ & $\uparrow$ & 15.50 \\
$\Xi_{c}(2815)^{0}(\frac{3}{2}^{-})$ & 2819.79  & $\uparrow$ & $\uparrow$ & 0.21 & $\Xi_{b}(6087)^{-}(?^{?})$  & 6087 & $\Xi_{b}(6084)\frac{1}{2}^{-}$ & 6084 & -3.00 \\
$\Xi_{c}(2882)^{0}(?^{?})$ & 2882  & $\Xi'_{c}(2896)\frac{1}{2}^{-}$ & 2896 & 14.00 & $\Xi_{b}(6095)^{0}(?^{?})$  & 6095.1 & $\Xi_{b}(6097)\frac{3}{2}^{-}$ & 6097 & 1.90 \\
$\Xi_{c}(2923)^{+}(?^{?})$ & 2922.8  & $\Xi'_{c}(2929)\frac{1}{2}^{-}$ & 2919 & -3.80 & $\Xi_{b}(6100)^{-}(\frac{3}{2}^{-})$  & 6099.8 & $\uparrow$ & $\uparrow$ & -2.80 \\
$\Xi_{c}(2923)^{0}(?^{?})$ & 2923.2  & $\uparrow$ & $\uparrow$ & -4.20 & $\Xi_{b}(6227)^{-}(?^{?})$  & 6227.9 & $\Xi_{b}(6224)\frac{1}{2}^{+}$ & 6224 & -3.90 \\
$\Xi_{c}(2930)^{+}(?^{?})$ & 2937.5  & $\Xi'_{c}(2934)\frac{3}{2}^{-}$ & 2934 & -3.50 & $\Xi_{b}(6227)^{0}(?^{?})$  & 6226.8 & $\uparrow$ & $\uparrow$ & -2.80 \\
$\Xi_{c}(2930)^{0}(?^{?})$ & 2938.55  & $\uparrow$ & $\uparrow$ & -4.55 & $\Xi_{b}(6327)^{0}(?^{?})$ & 6327.28 & $\Xi_{b}(6318)\frac{3}{2}^{+}$ & 6318 & -9.28 \\
$\Xi_{c}(2970)^{+}(\frac{1}{2}^{+})$ & 2964.3  & $\Xi_{c}(2949)\frac{1}{2}^{+}$ & 2949 & -15.30 & $\Xi_{b}(6333)^{0}(?^{?})$  & 6332.69 & $\Xi_{b}(6328)\frac{5}{2}^{+}$ & 6328 & -4.69 \\
$\Xi_{c}(2970)^{0}(\frac{1}{2}^{+})$ & 2965.9  & $\uparrow$ & $\uparrow$ & -16.90 & $\Omega_{b}^{-}(\frac{1}{2}^{+})$  & 6045.8 & $\Omega_{b}(6043)\frac{1}{2}^{+}$ & 6043 & -2.80 \\
$\Xi_{c}(3055)^{+}(\frac{3}{2}^{+})$ & 3055.9  & $\Xi_{c}(3061)\frac{3}{2}^{+}$ & 3061 & 5.10 & $\Omega_{b}(6316)^{-}(?^{?})$  & 6315.6 & $\Omega_{b}(6315)\frac{1}{2}^{-}$ & 6315 & -0.60 \\
$\Xi_{c}(3080)^{+}(\frac{5}{2}^{+})$ & 3077.2  & $\Xi_{c}(3078)\frac{5}{2}^{+}$ & 3078 & 0.80 & $\Omega_{b}(6330)^{-}(?^{?})$ & 6330.3 & $\Omega_{b}(6325)\frac{3}{2}^{-}$ & 6325 & -5.30 \\
$\Xi_{c}(3080)^{0}(\frac{5}{2}^{+})$ & 3079.9  & $\uparrow$ & $\uparrow$ & -1.90 & $\Omega_{b}(6340)^{-}(?^{?})$  & 6339.7 & $\Omega_{b}(6335)\frac{3}{2}^{-}$ & 6335 & -4.70 \\
$\Omega_{c}^{0}(\frac{1}{2}^{+})$ & 2695.3  & $\Omega_{c}(2696)\frac{1}{2}^{+}$ & 2696 & 0.70 & $\Omega_{b}(6350)^{-}(?^{?})$  & 6349.8 & $\Omega_{b}(6353)\frac{5}{2}^{-}$ & 6353 & 3.20 \\
\end{tabular}
\end{ruledtabular}
\end{table*}

\begin{table*}[htbp]
\begin{ruledtabular}\caption{The residual ($M_{cal.}$-$M_{exp.}$) plots of the 74 baryons in the $nL$-wave state groups for different families. The data are from Table~\ref{tb4}. $s$ denotes the number of strange quarks in baryons. $1D_{\lambda}$, $1P_{\lambda}$ and $1P_{\rho}$ denote $(0,2)1D$, $(0,1)1P$ and $(1,0)1P$, respectively. The subscripts $\lambda$ and $\rho$ represent the orbital excitation modes.}
\begin{tabular}{c |c | c | c | c | c | c }
\label{tb11}
         & $1S$  & $2S$ & $3S$ & $1D_{\lambda}$  & $1P_{\lambda}$ & $1P_{\rho}$  \\ \hline
$\Lambda_{c}(s=0)$& 1.54  & -2.60  &   & 16.90,~10.38 & 4.75,~3.00 & \\ \hline
$\Lambda_{b}(s=0)$& 2.43  &       &   & -11.2,~-5.60   & -13.2,~-7.07  & \\ \hline
$\Xi_{c}(s=1)$& 11.05, 8.56 & -15.30,  -16.90&   &   5.10, 0.80, -1.90& -2.90,~-4.90,~3.49,~0.21& \\ \hline
$\Xi_{b}(s=1)$&  9.00,~14.30   &  -3.90,~-2.80 &   & -9.28,~-4.69& -3.00,~1.90,~-2.80 & \\ \hline
$\Sigma_{c}(s=0)$ & 2.03,~3.35,~2.25,~15.58,~16.60,~15.52& ~-1.00     &   &    & -2.00,~7.00,~-7.00  &-9.00\\  \hline
$\Sigma_{b}(s=0)$ & 10.44,~5.36,~19.68,~15.26 &      &   &    &0.20,~-2.00 & \\ \hline
$\Xi'_{c}(s=1)$& 10.80,~10.30,~14.90,~14.30  &     &   &    &14.00,~-3.80,~-4.20,~-3.50,~-4.55 & \\ \hline
$\Xi'_{b}(s=1)$& 9.10,~18.70,~15.50   &     &   &    &  & \\ \hline
$\Omega_{c}(s=2)$&  0.70,~-1  &12.00     & -2.10  &    & 0.54,~2.83,~4.85 & -3.58,~-6.98\\ \hline
$\Omega_{b}(s=2)$&  -2.80  &      &    &    &-0.60,~-5.30,~-4.70,~3.20 &  \\
\end{tabular}
\end{ruledtabular}
\end{table*}

\end{document}